\documentclass[12pt,a4paper]{article}
\usepackage[latin2]{inputenc}
\usepackage[T1]{fontenc}
\usepackage{graphicx}
\usepackage{subfigure}
\usepackage{amsfonts}
\usepackage{amssymb}
\usepackage[intlimits]{amsmath}
\usepackage[mathscr]{eucal}
\usepackage{indentfirst}
\usepackage{bbm}
\usepackage{verbatim}
\makeatletter

\newcommand{\bigx}[1]{\bBigg@{#1}}
\newcommand{\p}{\partial}

\makeatother

\numberwithin{equation}{section}

\begin{document}

\title{{\normalsize\begin{flushright}ITP Budapest Report
    No. 625 \end{flushright}}\vspace{1cm}Boundary renormalisation group flows
    of unitary superconformal minimal models}

\author{M\'arton Kormos\footnote{kormos@general.elte.hu
}\\
\emph{\small Institute for Theoretical Physics, E\"otv\"os University, }\\
\emph{\small 1117 Budapest, P\'azm\'any P\'eter s\'et\'any 1/A, Hungary}\\
}

\maketitle
\thispagestyle{empty}

\begin{abstract}
In this paper we investigate renormalisation group flows of
supersymmetric minimal models generated by the boundary perturbing
field \(\hat G_{-1/2}\phi_{1,3}\). Performing the Truncated Conformal
Space Approach analysis the emerging pattern of the flow structure is
consistent with the theoretical expectations. According to the
results, this pattern can be naturally extended to those cases for
which the existing predictions are uncertain.
\end{abstract}

\newpage

\begin{section}{Introduction}
Conformal field theories with boundary attracted much interest
recently, due to their relevance in condensed matter physics, e.\ g.\
in the Kondo problem \cite{saleur} and their applications in
describing D-branes in string theory \cite{openstrings,tachyons}. In
terms of string theory the renormalisation group flow generated by a
boundary perturbing field corresponds to tachyon condensation and
exploring these flows can help in understanding the decay of
D-branes.

Many papers appeared in the literature about the boundary
perturbations and the corresponding renormalisation flows of unitary
minimal models \cite{reckrogg, gfunc, lesage, c=1,
grahamrunkelwatts}. Up to now, a systematic charting of the boundary
flows of the unitary \textsl{superconformal} minimal models has been
missing. Although there may be flows within the perturbative domain,
for a general study a nonperturbative tool is necessary. We choose the
Truncated Conformal Space Approach (TCSA), originally proposed in the
paper \cite{yurovzam} and applied to boundary problems in
\cite{doreypocklington...} and \cite{c=1}. The essence of the TCSA is
to diagonalise the Hamiltonian of the system on a subspace of the
infinite dimensional Hilbert space.

The paper is organised as follows. In section 2 we recall the basic
facts that we need about superconformal minimal models, including the
commutation rules of various fields. After briefly summarizing in
section 3 some main properties of these models in the presence of
boundaries, we turn to determining the Hamiltonian of the system and
defining the boundary flows in section 4. We also motivate our choice
for the perturbing operator in this section. Section 5 can be regarded
as the main part of the paper: we introduce the TCSA method and give
the details of the calculation of matrix elements. The theoretical
expectations are briefly summarized in section 6, while our results
are listed and compared to these expectations in section 7. Finally,
in section 8 we summarize the conclusions of the paper. Some numerical
data in tables and figures can be found in the Appendix.

\end{section}

\newpage

\begin{section}{Superconformal minimal models}
\begin{subsection}{The \(N=1\) superconformal algebra}
The algebra of superconformal transformations in the plane are
generated by two fields, \(T(z)\) and \(G(z)\) whose mode expansions
read
\begin{subequations}
\begin{align}
T(z)&=\sum_n L_n z^{-n-2}\,,\label{mode1}\\
G(z)&=\sum_r G_r z^{-r-3/2}\,.\label{mode2}
\end{align}
\label{modee}
\end{subequations}
\(T(z)\) is a spin 2 field, while \(G(z)\) has spin 3/2. Because of its
fermionic character two kinds of boundary conditions are possible for
\(G(z)\):
\begin{align}
G(e^{2\pi i}z)&=G(z)&&\rightarrow\text{Neveu--Schwarz sector,}\\
G(e^{2\pi i}z)&=-G(z)&&\rightarrow\text{Ramond sector.}
\end{align}
The \(N=1\) supersymmetric extension of the Virasoro algebra is
defined by the following (anti)commutation relations:
\begin{subequations}
\begin{align}
[L_n,L_m]&=(n-m)L_{n+m}+\frac{c}{12}n(n^2-1)\delta_{n,-m}\,,\label{a1}\\
\{G_r,G_s\}&=2L_{r+s}+\frac{c}3\left(r^2-1/4\right)\delta_{r,-s}\,,\;\label{a2}\\
[L_n,G_r]&=\left(\frac{n}2-r\right)G_{n+r}\,.\label{a3}
\end{align}
\label{algebra}
\end{subequations}
Here \(n\), \(m\) denote integers, while according to the boundary
conditions for \(G(z)\) the indices \(r\), \(s\) can take half-integer
(Neveu--Schwarz sector) or integer (Ramond sector) values.

The highest weight representations of the algebra are defined by 
highest weight states \(|h\rangle\) satisfying
\begin{subequations}
\begin{align}
L_n|h\rangle&=0\qquad\qquad n>0\,,\label{hw1}\\
L_0|h\rangle&=h\,,\label{hw2}\\
G_r|h\rangle&=0\qquad\qquad r>0\,.\label{hw3}
\end{align}
\label{hw}
\end{subequations}
The Verma module is generated by operators having a nonpositive
index. The vectors
\[L_{n_1}\dots L_{n_k}G_{r_1}\dots G_{r_l}|h\rangle\,,\qquad
n_1\le\dots\le n_k<0\,,\;\;r_1<\dots<r_l\le0\]
constitute a basis for the Verma module. Note that the inequalities
for the \(r_i\)-s are strict since
\(G_r^2=L_{2r}-\frac{c}{12}\,\delta_{r,0}\) by equation \eqref{a2}.

The full operator algebra of the Neveu--Schwarz and Ramond primary
fields is nonlocal. However, there exist projections that give local
operator algebras. One of them is the so-called fermion model, in
which one keeps only the Neveu--Schwarz sector.

In this paper we consider the so-called spin model which is obtained
in the following way. In both sectors a fermion parity operator
\(\Gamma\) can be introduced with algebraic relations
\([\Gamma,L_n]=\{\Gamma,G_r\}=0\). Thus every level of a Verma module
consists of \(\Gamma=\pm1\) eigenstates. The projection onto the even
parity (\(\Gamma=+1\)) states in the Neveu--Schwarz sector and onto
either the even (\(\Gamma=+1\)) or the odd (\(\Gamma=-1\)) parity
states in the Ramond sector yields a well defined local
theory. \label{projection}
\end{subsection}

\begin{subsection}{The minimal series}
The irreducible highest weight representations are the quotients of
the Verma modules by their maximal invariant submodules. In this paper
we deal with the superconformal minimal models which contain only a
finite number of Verma modules. In these models the Verma modules turn
out to be highly reducible. The superconformal minimal models are
indexed by two positive integers, \(p\) and \(q\) (we choose the
convention \(p<q\)). Their central charge is given by
\begin{equation}
c(p,q)=\frac32\left(1-2\frac{(p-q)^2}{pq}\right),\qquad\qquad \left(p\,,\frac{q-p}2\right)=1\,.
\end{equation}
Within a superconformal minimal model the highest weights are also
characterized by two integers, \(r\) and \(s\):
\begin{equation}
h(r,s)=\frac{(ps-qr)^2-(p-q)^2}{8pq}+\frac{1-(-1)^{r+s}}{32},\qquad
1\le r\le p-1,\;\;1\le s\le q-1\,.
\end{equation}
The sum \(r+s\) is even for a Neveu--Schwarz state and odd for a Ramond
state. Setting \(q=p+2\) gives the one parameter family of {\em
unitary} superconformal minimal models.

The fusion rule for the supersymmetric minimal models is 
\begin{equation}
\phi_{(r_1,s_1)}\otimes\phi_{(r_2,s_2)}=\sideset{}{'}\sum_{r'=|r_1-r_2|+1}^{\min\binom{2p-r_1-r_2-1}{r_1+r_2-1}}\sideset{}{'}\sum_{s'=|s_1-s_2|+1}^{\min\binom{2q-s_1-s_2-1}{s_1+s_2-1}}\phi_{(r',s')}\,,
\label{fusion}
\end{equation}
where the prime on the sums means that \(r'\) and \(s'\) jump by steps
of 2. Since no field can appear on the right hand side more than once,
the fusion rule coefficients are 1 for the fields appearing on the
right hand side and 0 otherwise.

\end{subsection}

\begin{subsection}{Field representations}
Since we will be interested in correlation functions, we have to
investigate how fields transform. Our discussion follows reference
\cite{watts}.  We denote the field corresponding to the state
\(|h\rangle\) by \(\phi_h(z)\). By definition
\(\phi_h(z)|0\rangle=\exp(zL_{-1})|h\rangle\), where \(|0\rangle\) is
the superconform invariant vacuum state.

\begin{subsubsection}{Virasoro properties}
\label{virprop}
The mode expansion \eqref{mode1} and the defining relations
(\ref{hw1}, \ref{hw2}) lead to the following operator product
expansion:
\begin{equation}
T(z)\phi(w)=\frac{h\phi(w)}{(z-w)^2}+\frac{\p \phi(w)}{z-w}+\dots
\end{equation}
Using standard contour integration technique one obtains the
commutation relation
\begin{equation}
L_n\phi(z)=z^n[(n+1)h+z\p]\phi(z)+\phi(z)L_n\,.
\end{equation}
Combining the relations for \(n=m\) and \(n=0\) one arrives at
\begin{equation}
[L_m-z^mL_0,\phi(z)]=hmz^m\phi(z)\,.
\label{Lorig}
\end{equation}

\end{subsubsection}

\begin{subsubsection}{Neveu--Schwarz case}
From equations \eqref{mode2}, \eqref{hw3} and \eqref{a2} we conclude
\begin{subequations}
\begin{align}
G(z)\phi(w)&=\frac{(\hat G_{-1/2}\phi)(w)}{z-w}+\dots\,,\label{gfi}\\
G(z)(\hat G_{-1/2}\phi)(w)&=\frac{2h\phi(w)}{(z-w)^2}+\frac{\p
  \phi(w)}{z-w}+\dots\,,
\end{align}
\end{subequations}
where \(\lim_{z\to0}(\hat G_{-1/2}\phi)(z)=G_{-1/2}|h\rangle\).

The commutation relations in this case turn out to be
\begin{subequations}
\begin{align}
G_r\phi(z)&=z^{r+1/2}(\hat G_{-1/2}\phi)(z)+\eta_\phi\phi(z)G_r\,,\label{G1strule}\\
G_r(\hat
G_{-1/2}\phi)(z)&=z^{r-1/2}((2r+1)h+z\p)\phi(z)-\eta_\phi(\hat G_{-1/2}\phi)(z)G_r\label{grgfi}
\end{align}
\label{Greduce}
\end{subequations}
with \(\eta_\phi=\pm1\) for fields of even and odd fermion number,
respectively.

\end{subsubsection}

\begin{subsubsection}{Ramond case}
There is a cut in the operator product of \(G(z)\) and a Ramond field
and thus a Ramond field changes the moding of \(G(z)\) from integral
to half-integral and vice versa. So it is impossible to obtain a
commutation relation of \(G_r\) with a Ramond field for fixed
\(r\). Fortunately, we can avoid using any (anti)commutation relation
for Ramond fields in our calculations. All we need is formally
generalizing relations \eqref{Greduce} for integral \(r\).  Then, just
like in the Virasoro case in section \ref{virprop}, combining the
\eqref{grgfi} equations for \(r=m\) and \(r=0\) yields
\begin{multline}
G_m(\hat G_{-1/2}\phi)(z)=z^mG_0(\hat G_{-1/2}\phi)(z)+2mhz^{m-1/2}\phi(z)\\
+\eta_\phi(\hat G_{-1/2}\phi)(z)(z^mG_0-G_m)\label{G2ndrule}\,.
\end{multline}
We will come back to these issues later in section \ref{ramondcase}.

\end{subsubsection}
\end{subsection}
\end{section}

\begin{section}{Superconformal field theory on the upper half plane}
The upper half plane (UHP) is the prototype of geometries having
nontrivial boundaries. The superconformal transformations then have
two roles. Firstly they connect the correlation functions on general
geometries to those on the UHP. On the other hand the transformations
that leave the geometry invariant put constraints on the correlation
functions on the UHP.

In the following we give a lightning review of the superconformal
boundary conditions and consistent boundary states based on paper
\cite{nepomechie}.

The derivation of the Ward identity on the UHP leads to a new
constraint on the operators of the chiral algebra. For the ordinary
stress energy tensor one obtains \(T_{xy}(x,0)=0\), which means that
there is no energy flow across the boundary (the real axis). In
complex coordinates this translates to the condition
\begin{equation}
T(z,\bar z)=\bar T(\bar z,z)\qquad \text{for Re}\,z=0.
\end{equation}
A strip of width \(L\) with boundary conditions \(\alpha\) and
\(\beta\) on the left and the right boundary, respectively, can be
mapped to the UHP by the exponential mapping, under which the
boundaries of the strip are mapped to the negative and positive part
of the real axis. The boundary conditions on the strip become
conformal boundary conditions on the plane. The discontinuity of the
boundary condition along the real axis can be described by a boundary
condition changing field \(\psi^{ab}\). These fields are in one to one
correspondence with the operator content of the theory on the strip.

Looking at the strip from the side (``closed channel'') and letting the
time flow in the direction of \(L\) the boundary conditions turn into
initial and final boundary states. Then the conformal invariance of
the boundary conditions translate to the following condition on these
boundary states:
\begin{equation}
(L_n-\bar L_{-n})|B\rangle=0\qquad\quad (B=\alpha,\;\beta)\,.
\label{llbar}
\end{equation}
For a
general chiral field \(W\) of spin \(s\) one finds
\begin{equation}
(W_n-(-1)^s\bar W_{-n})|B\rangle=0\,,
\end{equation}
which in our case means
\begin{equation}
(G_r\pm i\bar G_{-r})|B\rangle=0\,.
\end{equation}
This implies that the left-chiral and right-chiral part of the theory
are linked together. Only one copy of the supersymmetric Virasoro
algebra remains and all the calculations are chiral. It is also
necessary to carry out the projection described in section
\ref{projection}.

The equations for the state \(|B\rangle\) are solved in terms of the
so-called Ishibashi states \cite{ishibashi}, whose linear combinations
give the physically consistent boundary states. The further
constraints for these states, generalizations of the Cardy-equations
\cite{cardy}, come from the modular properties of the partition
function.

Now let us list the main results of this analysis. In the Ramond
sector the consistent boundary states are in one-to-one correspondence
with the irreducible highest weight representations of the algebra, so
they can be labeled by the same index set: we will denote them by
\(|\mathbf{k}^\text{R}\rangle\). In the Neveu--Schwarz sector there
are two physical boundary states for every irreducible representation,
they are denoted by \(|\mathbf{k}^\text{NS}\rangle\) and
\(|\mathbf{k}^{\widetilde{\mathrm{NS}}}\rangle\).

For the SM\((p,p+2)\) models (with \(p\) odd) the state content of the
Hilbert space of the theory on the strip with boundary conditions
\(\alpha\) and \(\beta\) is given by
\begin{subequations}
\begin{align}
\mathrm{tr}_\text{NS}\,e^{-RH_{\alpha\beta}}&=\sum_{i\in
\Delta_\text{NS}}n^i_{\alpha\beta}\chi^\text{NS}_i(q)\,,\\
\mathrm{tr}_\text{NS}\,\Gamma e^{-RH_{\alpha\beta}}&=\sum_{i\in
\Delta_\text{NS}} \widetilde{n}^i_{\alpha\beta}
\chi^{\widetilde{\mathrm{NS}}}_i(q)\,,\\
\mathrm{tr}_\text{R}\,e^{-RH_{\alpha\beta}}&=\sum_{i\in
\Delta_\text{R}}m^i_{\alpha\beta}\chi^\text{R}_i(q)\,,\\
\mathrm{tr}_\text{R}\,\Gamma e^{-RH_{\alpha\beta}}&=0\,,
\end{align}
\label{open}
\end{subequations}
where \(q=e^{-\pi R/L}\) and the characters are
\begin{subequations}
\begin{align}
\chi_i^\text{NS}(q)&=\mathrm{tr}_i\,q^{L_0-c/24}\,,& i&\in\Delta_\text{NS}\,,\\
\chi_i^{\widetilde{\mathrm{NS}}}(q)&=\mathrm{tr}_i\Gamma\,q^{L_0-c/24}\,,& i&\in\Delta_\text{NS}\,,\\
\chi_i^\text{R}(q)&=\mathrm{tr}_i\,q^{L_0-c/24}\,,& i&\in\Delta_\text{R}\,.
\end{align}
\end{subequations}
Using a generalised Verlinde formula the coefficients \(n^i_{\alpha\beta}\) and
\(m^i_{\alpha\beta}\) can be identified with the fusion rule
coefficients in the fusion
\(\Phi_\alpha\times\Phi_\beta\to\Phi_i\). Particularly,
\begin{subequations}
\begin{align}
n^i_{\mathbf{0}_\text{NS}\mathbf{k}_\text{NS}}&=\widetilde{n}^i_{\mathbf{0}_\text{NS}\mathbf{k}_\text{NS}}=\delta^i_k\,,
&m^i_{\mathbf{0}_\text{NS}\mathbf{k}_\text{NS}}&=0\,,\\
n^i_{\mathbf{0}_\text{NS}\mathbf{k}_{\widetilde{\mathrm{NS}}}}&=
-\widetilde{n}^i_{\mathbf{0}_\text{NS}\mathbf{k}_{\widetilde{\mathrm{NS}}}}=
\delta^i_k\,,&m^i_{\mathbf{0}_\text{NS}\mathbf{k}_{\widetilde{\mathrm{NS}}}}&=0\,,\\
n^i_{\mathbf{0}_\text{NS}\mathbf{k}_\text{R}}&=\widetilde{n}^i_{\mathbf{0}_\text{NS}\mathbf{k}_\text{R}}=0\,,&
m^i_{\mathbf{0}_\text{NS}\mathbf{k}_\text{R}}&=2\delta^i_k\,.
\end{align}
\label{n01}
\end{subequations}

The field content of the (projected) theory is thus given by
\begin{multline}
Z=\mathrm{tr}_\text{NS}\frac12(1+\Gamma)\,e^{-RH_{\alpha\beta}}+\mathrm{tr}_\text{R}\frac12(1+\Gamma)\,e^{-RH_{\alpha\beta}}=\\
\frac12\sum_{i\in
  \Delta_\text{NS}}\left(n^i_{\alpha\beta}\chi^\text{NS}_i(q)+\widetilde{n}^i_{\alpha\beta}\chi^{\widetilde{\mathrm{NS}}}_i(q)\right)+\frac12\sum_{i\in
  \Delta_\text{R}}m^i_{\alpha\beta}\chi^\text{R}_i(q)\,.
\end{multline}
If we choose \(\beta\) corresponding to
\(|\mathbf{0}^\text{NS}\rangle\) and \(\alpha\) to
\(|\mathbf{k}^\text{NS}\rangle\) (or
\(|\mathbf{0}^{\widetilde{\text{NS}}}\rangle\) and
\(|\mathbf{k}^{\widetilde{\text{NS}}}\rangle\)) then the right hand
side becomes
\begin{equation}
Z=\frac12\left(\chi^\text{NS}_k(q)+\chi^{\widetilde{\mathrm{NS}}}_k(q))\right)=
\mathrm{tr}_k^\text{NS}\frac12(1+\Gamma)\,q^{L_0-c/24}\,.
\label{znssame}
\end{equation}
In case we choose \(\beta\) corresponding to
\(|\mathbf{0}^\text{NS}\rangle\) and \(\alpha\) to
\(|\mathbf{k}^{\widetilde{\text{NS}}}\rangle\) (or
\(|\mathbf{0}^{\widetilde{\text{NS}}}\rangle\) and
\(|\mathbf{k}^\text{NS}\rangle\)) we get
\begin{equation}
Z=\frac12\left(\chi^\text{NS}_k(q)-\chi^{\widetilde{\mathrm{NS}}}_k(q))\right)=
\mathrm{tr}_k^\text{NS}\frac12(1-\Gamma)\,q^{L_0-c/24}\,.
\label{znsdiff}
\end{equation}
Finally, if we choose \(\beta\) corresponding to
\(|\mathbf{0}^\text{NS}\rangle\)
(or \(|\mathbf{0}^{\widetilde{\text{NS}}}\rangle\)) and \(\alpha\) to
\(|\mathbf{k}^\text{R}\rangle\) the result is
\begin{equation}
Z=\chi^\text{R}_k(q)\,.
\label{zr}
\end{equation}
The case of even \(p\) is more complicated due to the special
supersymmetric representation (\(\frac{p}2,\frac{p+2}2\)).

\end{section}

\begin{section}{Boundary renormalisation group flows}
In this paper we examine RG flows of superconformal minimal models
with boundary, generated by a boundary field. Since no bulk perturbing
field is allowed and the boundary perturbation preserves
supersymmetry, the RG flow takes place in the space of possible
supersymmetric boundary conditions, in which the fixed points are the
superconformal ones.

To study flows starting from a Cardy boundary condition it is
convenient to choose the boundary condition on one of the edges of the
strip to correspond to the \(h=0\) primary field, that is
\(\beta\sim|\mathbf{0}^\text{NS}\rangle\) or
\(|\mathbf{0}^{\widetilde{\text{NS}}}\rangle\). Then \eqref{znssame}
implies that if \(\alpha\) and \(\beta\) are both NS or
\(\widetilde{\mathrm{NS}}\) type boundary conditions, the Hilbert
space contains only the bosonic half of a single Neveu--Schwarz
module, the one corresponding to \(\alpha\). Similarly, according to
\eqref{znsdiff}, if \(\alpha\) is a NS type and \(\beta\) is a
\(\widetilde{\mathrm{NS}}\) type boundary condition (or vice versa),
the Hilbert space consists of the fermionic half of a single
\(\alpha\) module. Finally, as \eqref{zr} shows, if \(\alpha\) is a R
type boundary condition, the Hilbert space contains a single Ramond
module.

At the endpoint of the flow the final boundary condition can be read
off in a similar way simply from the spectrum of the
Hamiltonian. Since the boundary condition \(\beta\) does not change
along the flow (it is not perturbed), if we start with the bosonic
half of a NS Verma module (NS--NS or
\(\widetilde{\mathrm{NS}}\)--\(\widetilde{\mathrm{NS}}\)) and at the
end of the flow we find bosonic part(s) of Verma module(s) then this
is a NS\(\to\)NS (or
\(\widetilde{\mathrm{NS}}\to\widetilde{\mathrm{NS}}\)) type
flow. Similarly, if we identify fermionic part(s) then this indicates
a NS\(\to\widetilde{\mathrm{NS}}\) (or
\(\widetilde{\mathrm{NS}}\to\)NS) type flow.

We have to choose the perturbing field on the boundary \(\alpha\). We
would like a perturbation that preserves supersymmetry
off-critically. Ramond operators do not spoil supersymmetry but they
are boundary condition changing operators, so they cannot live on a
boundary. Neveu--Schwarz fields of the form \((\hat
G_{-1/2}\phi_\text{pert})\) also preserve supersymmetry and they can
be located on the boundary, so we choose this kind of
perturbation. The perturbing operator should be relevant, that is its
scaling dimension must be less than one. Finally, since the
Hamiltonian is bosonic, \(\phi_\text{pert}\) must be fermionic. The
choice \(\phi_\text{pert}=\phi_{1,3}\) satisfy all these
conditions. \((\hat G_{-1/2}\phi_{1,3})\) has conformal weight
\(p/q<1\) and \(\phi_{1,3}\) is fermionic. The latter can be seen from
the fact that in the free fermion limit \(p,q\to\infty\)
(\(q-p=\text{const.}\)) \(h_{1,3}\to1/2\).

Our choice also has the advantage that this operator can live on every
boundary that has supersymmetric relevant perturbations. This can be
seen from the following. An operator \(\psi\) living on the boundary
of the strip is mapped under the exponential map to an operator living
on the real axis. We want this operator to leave the boundary
condition unchanged along the axis, which means that the coefficients
in \eqref{open} for \(\alpha=\beta\) should be nonzero. Since they can be related to the
fusion coefficients this means that the representation \(h_\psi\)
should appear in the fusion of \(\phi_\alpha\) with
itself\footnote{The coefficients \(\widetilde{n}^i_{\alpha\beta}\) are
not fusion coefficients themselves, but this does not change the
argument essentially.}. The field \(\phi_{1,3}\) appears in the self
fusion of all fields except for the ones in the first and the last
column of the superconformal Kac table (\(\phi_{(1,s)}\),
\(\phi_{(p-1,s)}\)), but the corresponding boundaries do not have any
relevant supersymmetric perturbations at all. Also note that being a
bosonic field, the fusion with \((\hat G_{-1/2}\phi_{1,3})\) maps the
projected subspaces onto themselves.

The perturbation \((\hat G_{-1/2}\phi_{1,3})\) is an integrable one
\cite{mathieu}, but we shall not use this property in our analysis.

The Hamiltonian of the perturbed superconformal field theory takes the
form
\begin{equation}
H=H_\text{CFT}+H_\text{pert}\,,
\end{equation}
where \(H_\text{CFT}\) is the Hamiltonian of the unperturbed theory
\begin{equation}
H_\text{CFT}=H_{\alpha0}=\frac\pi{L}(L_0-\frac{c}{24})
\end{equation}
and \(H_\text{pert}\) comes from the perturbation on the strip:
\begin{equation}
H_\text{pert}^\text{strip}=\lambda\,(\hat G_{-1/2}\phi_{1,3})(0,0)\,.
\end{equation}
The location of the left boundary is at \(x=0\) and we are free to
choose \(t=0\) for calculating the spectrum. By the exponential map
this on the \(z\)-plane becomes
\begin{equation}
H_\text{pert}=\lambda\,\left(\frac\pi{L}\right)^{h_{1,3}+1/2}(\hat G_{-1/2}\phi_{1,3})(1)_z\,.
\end{equation}
Thus the complete Hamiltonian on the plane can be written as
\begin{equation}
H=\frac\pi{L}\left[L_0-\frac{c}{24}+\lambda\left(\frac{L}\pi\right)^{1/2-h_{1,3}}(\hat{G}_{-1/2}\phi_{1,3})(1)\right]\,.
\end{equation}
Since during a numerical calculation one works with dimensionless
quantities we have to measure the volume (\(L\)) and the energies in
some typical mass or energy difference of the model, \(M\). In other
words, we use the dimensionless quantities \(l=ML\),
\(\varepsilon=E/M\), \(\kappa=\lambda/M^{1/2-h_{1,3}}\) and \(h=H/M\): 
\begin{equation}
h=\frac\pi{l}\left[L_0-\frac{c}{24}+\kappa\left(\frac{l}\pi\right)^{1/2-h_{1,3}}(\hat{G}_{-1/2}\phi_{1,3})(1)\right]\,.
\end{equation}

The renormalisation group flow can be implemented by varying the
volume \(l\) while keeping the coupling constant \(\kappa\)
fixed. Equivalently -- and we choose this way -- one can keep \(l\)
fixed and vary \(\kappa\) on some interval. Starting from
\(\kappa=0\), which is the ultraviolet (UV) limit, the matrix \(h\)
can be diagonalised at different positive and negative values of
\(\kappa\). In both directions the flow approaches a fixed point, that
is a new supersymmetric conformal boundary condition. By \eqref{open}
we should observe the eigenstates rearranging themselves into a direct
sum of super Verma modules (see the figures in the Appendix).  Since
by \eqref{n01} the coefficients in this sum can only take values 0 and
1, we expect a simple sum of super Verma modules at the end of the
flows. From the degeneracy pattern of the eigenvalues we can identify
the modules (the boundary conditions) using the characters and weight
differences of the supersymmetric minimal model in question.

\end{section}

\begin{section}{The Truncated Conformal Space Approach}
The method we use for organising the boundary flows is the so-called
truncated conformal space approach, or TCSA. In this approach the
infinite dimensional Hilbert space is truncated to a finite
dimensional vector space by using only those states whose energy is
not greater than a threshold value, \(E_\text{cut}\). This is
equivalent to truncating the Hilbert space at a given level. The
Hamiltonian is then diagonalised on this truncated space. One can
think of this procedure as being equivalent to the variational method:
the Ansatzes for the energy eigenstates of the perturbed Hamiltonian
are finite linear combinations of the eigenstates of the conformal
Hamiltonian.

It is important that the errors of the TCSA diagonalisation are not
under control. For example, it may happen that before the flow reaches
the scaling region the truncation errors start to dominate. If we use
various cuts and find that the flow picture does not change
drastically (only the precision of the result gets higher with higher
cuts), then it means that the unpleasant case mentioned above does not
happen.

Of course establishing the endpoint of the renormalisation group flow
by TCSA cannot be regarded as a conclusive proof. What one can see is
that the flow goes in the vicinity of some superconformal boundary
condition. The exact infrared fixed point can never be reached by TCSA
because of the truncation (this can be observed in figure
\ref{5,7,1,5}). We are looking for the range where the TCSA trajectory
is closest to the fixed point. 

\begin{subsection}{The matrix elements of the Hamiltonian}
There are two ways for determining the eigenvalues of the total
Hamiltonian. The first is to use an orhonormal basis for calculating
the matrix elements, which requires an orthogonalisation process. We
can avoid this by chosing the second way, in which the basis is not
orhogonal. But then the numerical matrix to be diagonalised is
\begin{equation}
h_{ij}=\langle f_i|h|e_j\rangle\,,
\end{equation}
where \(\{f_i\}\) are the elements of the reciprocal basis, that is
\begin{equation}
\langle f_i|e_j\rangle=\delta_{ij}\,.
\end{equation}
This amounts to calculating the matrix element
\begin{equation}
h_{ij}=(M^{-1})_{ik}\langle e_k|h|e_j\rangle\,,
\end{equation}
where \(M\) is the inner product matrix:
\begin{equation}
M_{ij}=\langle e_i|e_j\rangle\,.
\end{equation}
Using the notation
\begin{equation}
B_{ij}=\langle e_i|(\hat G_{-1/2}\phi_{1,3})(1)|e_j\rangle
\end{equation}
it can be expressed as
\begin{equation}
h_{ij}=\frac\pi{l}\left[(h_i-\frac{c}{24})\delta_{ij}+\kappa\left(\frac{l}\pi\right)^{1/2-h_{1,3}}(M^{-1}B)_{ij}\right].
\end{equation}
Here we made use of the fact that the basis elements \(\{e_i\}\) are
eigenvectors of \(H_0\). 
\end{subsection}

\begin{subsection}{Computing the matrix elements of the perturbing
    field between descendant states}
The task is now to calculate the matrices \(M\) and \(B\) and then to
determine the eigenvalues of the resulting matrix \(h\). The Hilbert
space consists of the Verma module of \(\phi_\alpha\) (see eq.\
\eqref{open}) and as mentioned above, we choose the following basis
elements:
\begin{equation}
L_{n_1}\dots L_{n_k}G_{r_1}\dots G_{r_l}|h\rangle\,,\qquad
n_1\le\dots\le n_k<0\,,\;\;r_1<\dots<r_l\le0\,.
\end{equation}
Due to the singular vectors, not all of these are linearly
independent. One can obtain a basis for the irreducible submodule by
requiring the nonsingularity of the inner product matrix level by
level. In the Ramond sector we kept the states which had even number
of \(G_r\) operators. Chosing the odd parity sector instead gives
exactly the same result: the spectrum of the Hamiltonian falls into
two identical copies.

The calculation of \(h\) is now a question of algebraic
manipulations. We use the basic algebraic relations \eqref{algebra},
\eqref{hw} and the definition of adjoint operators
\begin{align}
L_n^\dagger&=L_{-n}\,,\\
G_n^\dagger&=G_{-n}\,.
\end{align}
In the Ramond case during the calculation of the inner product matrix
\(M\) we also demand that \(\langle\alpha|G_0|\beta\rangle=0\)
(\(\alpha\) and \(\beta\) are primary), because the inner product of
states with opposite \(\Gamma\)-parity must vanish. 

The calculation of the matrix \(B\) is more difficult because it
requires the usage of some commutation rules for reducing the matrix
elements. However, these commutation rules are different for the
Ramond and the Neveu--Schwarz case. Let us deal with the two cases in
turn.

\begin{subsubsection}{Ramond case}
\label{ramondcase}
In the Ramond case we can use the already mentioned commutational
rules \eqref{Lorig}, \eqref{G1strule} and \eqref{G2ndrule}. We recall
them because they lie at the heart of the algorythm. For the modes
\(L_n\) we saw (cf.\ \eqref{Lorig})
\begin{subequations}
\begin{align}
L_n\phi(z)&=z^nL_0\phi(z)+\phi(z)(nhz^n-z^nL_0+L_n)\,,\label{L1strule}\\
\phi(z)L_n&=z^n\phi(z)L_0+(-nhz^n-z^nL_0+L_n)\phi(z)\,.
\label{Lreduce}
\end{align}
\end{subequations}
Similarly
\begin{subequations}
\begin{multline}
L_n(\hat G_{-1/2}\phi)(z)=z^nL_0(\hat G_{-1/2}\phi)(z)\\
+(\hat G_{-1/2}\phi)(z)(n(h+1/2)z^n-z^nL_0+L_n)\,,
\end{multline}\\
\vspace{-1.5cm}
\begin{multline}
(\hat G_{-1/2}\phi)(z)L_n=z^n(\hat G_{-1/2}\phi)(z)L_0\\
+(-n(h+1/2)z^n-z^nL_0+L_n)(\hat G_{-1/2}\phi)(z)\,.
\end{multline}
\label{Lrules}
\end{subequations}
We make use of these formulae in reducing the three point
functions. Here is an example of the application of rule
\eqref{L1strule}:
\begin{multline}
\langle L_{-n}\mathcal{O}_1\alpha|\phi_{1,3}(1)|\mathcal{O}_2\alpha\rangle= \langle\mathcal{O}_1
\alpha|L_n\phi_{1,3}(1)|\mathcal{O}_2\alpha\rangle\\
=(h_\alpha+nh_{1,3}-h_\alpha)\langle
\mathcal{O}_1\alpha|\phi_{1,3}(1)|\mathcal{O}_2\alpha\rangle+\langle
\mathcal{O}_1\alpha|\phi_{1,3}(1)|L_n\mathcal{O}_2\alpha\rangle\,,
\label{Lexample}
\end{multline}
where \(\mathcal{O}_1\) and \(\mathcal{O}_2\) are arbitrary strings of
lowering operators.

The counterpart of these relations for the modes \(G_r\) are (see
\eqref{G1strule})
\begin{subequations}
\begin{align}
G_r\phi(z)&=z^{r+1/2}(\hat G_{-1/2}\phi)(z)+\eta_\phi\phi(z)G_r\,,\label{Grule1}\\
\phi(z)G_r&=-\eta_\phi z^{r+1/2}(\hat G_{-1/2}\phi)(z)+\eta_\phi G_r\phi(z)\,,
\end{align}
\label{Grules1}
\end{subequations}
and (cf.\ \eqref{G2ndrule})
\begin{subequations}
\begin{multline}
G_m(\hat G_{-1/2}\phi)(z)=z^mG_0(\hat G_{-1/2}\phi)(z)+2mhz^{m-1/2}\phi(z)\\
+\eta_\phi(\hat G_{-1/2}\phi)(z)(z^mG_0-G_m)\,,
\end{multline}\\
\vspace{-1.5cm}
\begin{multline}
(\hat G_{-1/2}\phi)(z)G_m=z^m(\hat G_{-1/2}\phi)(z)G_0+\eta_\phi2mhz^{m-1/2}\phi(z)\\
+\eta_\phi(z^mG_0-G_m)(\hat G_{-1/2}\phi)(z)\,.
\end{multline}
\label{Grules2}
\end{subequations}

These rules can be used in a similar way to \eqref{Lexample}, but only if
\(\phi_\alpha\) is a Ramond field (\(G_0\phi_\alpha\) is not defined
for Neveu--Schwarz fields). Iteratively applying the rules
\eqref{Lrules} and \eqref{Grules1}, \eqref{Grules2} to a matrix
element leads to a linear combination of the following correlation
functions:
\begin{subequations}
\begin{align}
\langle\alpha&|(\hat G_{-1/2}\phi_{1,3})(1)|\alpha\rangle,\label{res1}\\
\langle\alpha&|\phi_{1,3}(1)|\alpha\rangle,\label{res2}\\
\langle G_0\alpha&|\phi_{1,3}(1)|\alpha\rangle,\\
\langle \alpha&|\phi_{1,3}(1)|G_0\alpha\rangle.
\end{align}
\end{subequations}
The second of these always equals to zero in our case, because
\(\phi_{1,3}\) is a fermion-like operator. This also
means that \(\eta_{\phi_{1,3}}=-1\).

Due to \eqref{Grule1}
\begin{equation}
\langle G_0\alpha|\phi_{1,3}(1)|\alpha\rangle= \langle\alpha|(\hat
G_{-1/2}\phi_{1,3})(1)|\alpha\rangle+\eta_\phi \langle
\alpha|\phi_{1,3}(1)|G_0\alpha\rangle.
\end{equation}
Since \(G_0^\dagger=G_0\)
\begin{equation}
\langle G_0\alpha|\phi_{1,3}(1)|\alpha\rangle=\langle \alpha|\phi_{1,3}(1)|G_0\alpha\rangle,
\end{equation}
which implies for \(\eta_{\phi_{1,3}}=-1\)
\begin{equation}
\langle G_0\alpha|\phi_{1,3}(1)|\alpha\rangle=\frac12\langle\alpha|(\hat G_{-1/2}\phi_{1,3})(1)|\alpha\rangle.
\end{equation}
We see that every matrix element can be traced back to the three point
function
\begin{equation}
\langle\alpha|(\hat G_{-1/2}\phi_{1,3})(1)|\alpha\rangle=\langle\phi_\alpha(\infty),(\hat G_{-1/2}\phi_{1,3})(1),\phi_\alpha(0)\rangle
\end{equation}
which simply equals to a structure constant. However, we do not need
the numerical value of this constant, because it can be absorbed into
the coupling constant \(\kappa\). The physical scale is fixed by the
value of \(\kappa\) where the crossover takes place. Since we are
interested only in the asymptotic behaviour of the flows (the fixed
points), we do not need to fix the real physical scale.

\end{subsubsection}

\begin{subsubsection}{Neveu--Schwarz case}
In the Neveu--Schwarz case we can derive the rules for reducing the
three point functions using contour integration technique. From
\begin{multline}
\langle L_nA(\infty)|B(0)\rangle=\langle
A(\infty)|L_{-n}B(0)\rangle=\langle A(\infty)|\oint_0\frac{dz}{2\pi
i}z^{-n+1}T(z)B(0)\rangle\\
=\langle-\oint_\infty\frac{dz}{2\pi i}z^{-n+1}T(z)A(\infty)|B(0)\rangle
\end{multline}
we obtain
\begin{equation}
L_n\Phi(\infty)=-\oint_\infty\frac{dz}{2\pi i}z^{-n+1}T(z)\Phi(\infty)\,.
\end{equation}
Then for a three point function:
\begin{multline}
\langle L_{-n}A(\infty)|B(1)|C(0)\rangle=-\oint_\infty\frac{dz}{2\pi i}z^{n+1}\langle T(z)A(\infty)|B(1)|C(0)\rangle\\
=\oint_1\frac{dz}{2\pi i}\sum_{k=-1}^n\binom{n+1}{k+1}(z-1)^{k+1}\langle
A(\infty)|T(z)B(1)|C(0)\rangle\\
+\oint_0\frac{dz}{2\pi i}z^{n+1}\langle A(\infty)|B(1)|T(z)C(0)\rangle\\
=\langle A(\infty)|B(1)|L_nC(0)\rangle+\sum_{k=-1}^n\binom{n+1}{k+1}\langle A(\infty)|L_k B(1)|C(0)\rangle\,.
\end{multline}
With similar manipulations one can obtain rules for the remaining two
initial positions of \(L_{-n}\) and their counterparts for the modes
\(G_{-r}\). These formulae can be obtained from the algebraic
commutation relations as well. Now they can be applied iteratively to
the three point function in the following manner.

First we throw down all operators from the first field to the others:
\begin{subequations}
\begin{align}
\langle L_{-n}A|B(1)|C\rangle&=\langle
A|B(1)|L_n C\rangle+\sum_{k=-1}^n\binom{n+1}{k+1}\langle A|L_k B(1)|C\rangle\,,\\
\langle G_{-n}A|B(1)|C\rangle&=\eta_B\langle
A|B(1)|G_n C\rangle+\sum_{k=-1/2}^n\binom{n+\frac12}{k+\frac12}\langle A|G_kB(1)|C\rangle\,.
\end{align}
\end{subequations}

Then, exploiting that \(A\) is now primary we throw everything from the
rightmost field to the middle one:
\begin{subequations}
\begin{align}
\langle A|B(1)|L_{-n}C\rangle&=-\sum_{k=-1}^\infty\binom{-n+1}{k+1}\langle A|L_kB(1)|C\rangle\,,\\
\langle A|B(1)|G_{-n}C\rangle&=-\eta_B\sum_{k=-1/2}^\infty\binom{-n+\frac12}{k+\frac12}\langle A|G_kB(1)|C\rangle\,.
\end{align}
\label{right}
\end{subequations}
Certainly, in practice the sums truncate at the level of
\(B\). Finally we eliminate the operators from the field \(B\) (\(A\)
and \(C\) are now primary):
\begin{subequations}
\begin{align}
\langle A|L_{-n}B(1)|C\rangle&=(-1)^n\langle A|B(1)|L_{-1}C\rangle+(-1)^n(n-1)\langle A|B(1)|L_0 C\rangle\,,\\
\langle A|G_{-n}B(1)|C\rangle&=\eta_B(-1)^{n+1/2}\langle A|B(1)|G_{-1/2}C\rangle\,.
\end{align}
\label{middle}
\end{subequations}
Now we can throw this \(L_{-1}\) and \(G_{-1/2}\) back onto the middle
field using \eqref{right} and repeat step \eqref{middle}. Eventually
some \(L_{-1}\) and possibly one \(G_{-1/2}\) can accumulate on the
middle field. The effect of operators \(L_{-1}\) is simply
differentiating. Since conformal symmetry determines the functional
form of the three point function of quasi-primary fields we obtain:
\begin{multline}
\langle A|L_{-1}^mB(1)|C\rangle=(h_A-h_B-h_C)(h_A-h_B-h_C-1)\dots\\
\dots(h_A-h_B-h_C-m+1)\langle A|B(1)|C\rangle\,.
\end{multline}
The final result is -- like in the Ramond case -- a linear
combination of matrix elements \eqref{res1}, \eqref{res2} where the
second one gives zero again.

The reason for not using contour integral technique in the Ramond
case is the presence of the cut in the operator product of \(G(z)\)
with a Ramond field. 

It is interesting to note that albeit \(G_0\phi_\alpha\) is not
really well-defined for \(\phi_\alpha\) being a Neveu--Schwarz field,
the Ramond method explained in section \ref{ramondcase} still works in
the Neveu--Schwarz case and gives the same result as the contour
integral method.

\end{subsubsection}

\end{subsection}

\end{section}

\begin{section}{Previous results and expectations}
Before turning to the result of the TCSA let us summarize the
theoretical predictions for the boundary flows found in the
literature. Our main resource is the paper of Fredenhagen
\cite{fredenhagen}, where the author gives some predictions for
boundary flows in general coset models. Here we only summarize the
consequences for the special case of superconformal minimal models.

The SM\((p,p+2)\) unitary superminimal models are realized by the 
\begin{equation}
\frac{su(2)_k\otimes su(2)_2}{su(2)_{k+2}}
\end{equation}
coset models with \(p=k+2\). 
Let us label the sectors of \(su(2)_k\) by \(l=0,1,\dots,k\), the
sectors of \(su(2)_2\) by \(m=0,1,2\) and the sectors of
\(su(2)_{k+2}\) by \(l'=0,1,\dots,k+2\). The sectors of the direct
product are thus labeled by the pairs \((l,m)\). These sectors split
up with respect to the \(su(2)_{k+2}\) subalgebra,
\begin{equation}
V_{l,m}=\bigoplus_{l'}W_{l,m,l'}\otimes V_{l'}\,,
\end{equation}
where \(W_{l,m,l'}\) are the sectors of the coset theory. Only those
coset sectors are allowed for which
\begin{equation}
l+m+l'\;\;\;\text{is even.}
\end{equation}
Furthermore, there are identifications between these admissible representations:
\begin{equation}
(l,m,l')\cong(k-l,2-m,k+2-l')\,.
\label{ident}
\end{equation}
Using the \(p=k+2\) rule and the new variables (the Kac-indices) 
\begin{subequations}
\begin{alignat*}{2}
r&=l+1&\qquad\qquad&r=1,2,\dots,p-1\,,\\
s&=l'+1&&s=1,2,\dots,p+1
\end{alignat*}
\end{subequations}
the precise relations with the superminimal fields are (see \cite{gko}):
\begin{align}
(r,s)_\text{NS}&=(r-1,0,s-1)\oplus(r-1,2,s-1)\,,\label{nsdirsum}\\
(r,s)_{\text{R}\;\,}&=(r-1,1,s-1)\,.
\end{align}
In the Neveu--Schwarz sector the direct sum in \eqref{nsdirsum}
corresponds to the sum of the \(\Gamma\)-projected subspaces. The
identification \eqref{ident} translates to
\begin{equation}
(r,s)\cong(p-r,p+2-s)\,,
\end{equation}
which is the well-known symmetry relation of the Kac table of the
superminimal models.

In the A-series of superminimal models the boundary conditions
\(\alpha\) are labeled by triplets \((l,m,l')\) taking values in the
same range as the sectors including selection and identification
rules. The NS and \(\widetilde{\mathrm{NS}}\) boundary conditions
correspond to the \(m=0,2\) choices. The agreement between our results
and the predictions for the flows justifies this correspondence. 

The boundary flows are predicted in the following way. First we have
to find a boundary condition \(\alpha\) and a representation \(S\) so
that
\begin{equation}
(0,S^+|_{su(2)_{k+2}})\times\alpha=(l,m,l')\,,\qquad\qquad(l+m+l'\;\text{
    is even})\,.
\end{equation}
Then a boundary flow is predicted between the following coset boundary
configurations:
\begin{equation}
X:=(0,S^+|_{su(2)_{k+2}})\times\alpha\longrightarrow(S,0)\times\alpha=:Y\,.
\end{equation}
If we want to study flows starting from the boundary condition
\((l,m,l')\) with \(0\le l'\le k\) we can set \(\alpha=(l,m,0)\) and
\(S=(l',0)\). This choice corresponds to the flow
\begin{equation}
(l,m,l')\longrightarrow\bigoplus_J N_{l,l'}^{(k)J} (J,m,0)\,,
\label{cosetIgen}
\end{equation}
where the \(N_{l,l'}^{(k)J}\) are the fusion coefficients in
\(su(2)_k\).\\
Similarly, in the case of \(2\le l'\le k+2\) the choice
\(\alpha=(k-l,2-m,0)\) and \(S=(k+2-l',0)\) leads to the flow
\begin{equation}
(l,m,l')\longrightarrow\bigoplus_J N_{l,l'-2}^{(k)J} (J,2-m,0)\,.
\end{equation}
The boundary conditions which are not covered by these rules
are\footnote{The author is thankful to S.\ Fredenhagen for the private
discussion on these cases.} for \(l'=k+1,\,1\). These correspond to
the \((r,p)\) and \((r,2)\) states. (The boundary conditions
\((r,p+1)\) and \((r,1)\) do not have relevant supersymmetric
perturbations.)
\newline
For \(l'=k+1\) the suitable choice is \(S=(0,1)\) and
\(\alpha=(l,m,k+2)\). Then
\begin{align}
(l,0\,\text{or}\,2,k+1)&\longrightarrow(l,1,k+2)\cong(k-l,1,0)\,,\\
(l,1,k+1)&\longrightarrow(k-l,0,0)\oplus(k-l,2,0)\,.
\end{align}
Finally, for \(l'=1\) one should choose \(S=(0,1)\) and
\(\alpha=(l,m,0)\), which yields the flows
\begin{align}
(l,0\,\text{or}\,2,1)&\longrightarrow(l,1,0)\,,\\
(l,1,1)&\longrightarrow(l,0,0)\oplus(l,2,0)\,.
\end{align}
Let us summarize the \(l=0\) case, when these results give
\begin{alignat}{2}
(0,m,l')\longrightarrow&&\;(l',m,0)&\quad\quad\qquad0\le l'\le k\,,\label{I.gen}\\
(0,0\,\text{or}\,2,k+1)\longrightarrow&&\;(k,1,0)&\quad\quad\qquad l'=k+1\,,\label{I.NS}\\
(0,1,k+1)\longrightarrow&&\;(k,0,0)\oplus(k,2,0)&\,,\label{I.R}\\
&&&\nonumber\\
(0,m,l')\longrightarrow&&\;(l'-2,2-m,0)&\quad\quad\qquad2\le l'\le  k+2\,,\label{II.gen}\\
(0,1,1)\longrightarrow&&\;(0,0,0)\oplus(0,2,0)&\,.\label{II.R}
\end{alignat}
Multiplying both sides of these rules by \((l,0,0)\) we get back the
rules for general \(l\). In the Kac index language this is the
,,disorder line rule'', discussed in paper \cite{disorder}: if there
is a flow between boundary conditions \(\alpha\) and \(\beta\), then
there exists a flow between \(\alpha\times\gamma\) and
\(\beta\times\gamma\) where \(\times\) denotes the fusion product. For
the boundary states of \(N=1\) superminimal models the fusion rule
\eqref{fusion} should be used together with the rules\footnote{These
are correct with the choice \(m=0\iff\mathrm{NS}\) and
\(m=2\iff\widetilde{\mathrm{NS}}\). In case of the other choice a
\(\mathrm{NS}\leftrightarrow\widetilde{\mathrm{NS}}\) swap is needed.}
\begin{subequations}
\begin{align}
\text{NS}\times\mathrm{NS}&=\widetilde{\mathrm{NS}}\times\widetilde{\mathrm{NS}}=\text{NS}\,,\\
\text{NS}\times\widetilde{\mathrm{NS}}&=\widetilde{\mathrm{NS}}\,,\\
\text{R}\times\text{NS}&=\text{R}\times\widetilde{\mathrm{NS}}=\text{R}\,,\\
\text{R}\times\text{R}&=\text{NS}\oplus\widetilde{\mathrm{NS}}\,.
\end{align}
\end{subequations}
This means that we only need to determine the flows starting
from states in the first row of the Kac table, all the other flows are
given by the fusion rules of the operator algebra.

The authors of paper \cite{ahn-rim} have also derived the flows for
some Neveu--Schwarz states in the first row of the Kac table, but only for even
\(p\).

\end{section}

\begin{section}{TCSA results}
We were searching the answers for the following questions.
\begin{enumerate}
\item Does our TCSA analysis affirm or disprove the predictions of the Fredenhagen rules?
\item Do all the flows obey the disorder line rule?
\item What happens in the \(p\) even models, where both these rules
  and the classification of consistent boundary conditions are
  somewhat certain?
\end{enumerate}

We have positive answers for the first two questions. Namely, the
following rules can be read out from the TCSA analysis of the boundary
flows of the unitary superconformal minimal model SM\((p,p+2)\)
perturbed by the operator \(\hat G_{-1/2}\phi_{1,3}\) (see also the
Appendix):
\label{class}
\begin{enumerate}
\item For \(\kappa>0\):

\begin{enumerate}
\item For \(2\le s\le p-1\)
\begin{subequations}
\begin{align}
(1,s)_\text{NS}&\longrightarrow(s,1)_\text{NS}\,,\\
(1,s)_{\widetilde{\mathrm{NS}}}&\longrightarrow(s,1)_{\widetilde{\mathrm{NS}}}\,,\\
(1,s)_{\text{R}\;\,}&\longrightarrow(s,1)_\text{R}\,.
\end{align}
\end{subequations}
This agrees with the Fredenhagen result \eqref{I.gen}. According
to the disorder line rule we found (cf.\ \eqref{cosetIgen}):
\begin{equation}
(r,s)\to(|s-r|+1,1)\oplus(|s-r|+3,1)\oplus\dots\oplus(\textstyle{\min\binom{s+r-1}{2p-(s+r)-1}},1)\,.
\label{rule1a}
\end{equation}
\label{comment}These flows do not change the type of the boundary condition so on
the right hand side all the boundary conditions are of the same
type. Thus, if two of them have different fermion parity, then the
half-integer levels of one should combine with the integer levels of
the other. That is exactly what we observed.
\item For \(s=p\)
\begin{subequations}
\begin{align}
(1,p)_\text{NS}&\longrightarrow(p-1,1)_\text{R}\,,\\
(1,p)_{\widetilde{\mathrm{NS}}}&\longrightarrow(p-1,1)_\text{R}\,,\\
(1,p)_{\text{R}\;\,}&\longrightarrow(p-1,1)_\text{NS}\oplus(p-1,1)_{\widetilde{\mathrm{NS}}}\,.
\end{align}
\end{subequations}
Note that if the initial boundary condition is of Neveu--Schwarz type,
then the result of the flow is a Ramond type boundary condition and
vice versa.  We also identified the flows related to the
flow \eqref{rule1b} by the disorder line rule:
\begin{subequations}
\begin{align}
(r,p)_\text{NS}&\longrightarrow(p-r,1)_\text{R}\,,\\
(r,p)_{\widetilde{\mathrm{NS}}}&\longrightarrow(p-r,1)_\text{R}\,,\\
(r,p)_{\text{R}\;\,}&\longrightarrow(p-r,1)_\text{NS}\oplus(p-r,1)_{\widetilde{\mathrm{NS}}}\,.
\end{align}
\label{rule1b}
\end{subequations}

\end{enumerate}
\item For \(\kappa<0\):
\begin{enumerate}
\item For \(3\le s\le p\)
\begin{subequations}
\begin{align}
(1,s)_\text{NS}&\longrightarrow(s-2,1)_{\widetilde{\mathrm{NS}}}\,,\\
(1,s)_{\widetilde{\mathrm{NS}}}&\longrightarrow(s-2,1)_\text{NS}\,,\\
(1,s)_{\text{R}\;\,}&\longrightarrow(s-2,1)_\text{R}\,.
\end{align}
\end{subequations}
This again gives back the Fredenhagen flow \eqref{II.gen}. We found
that the disorder line rule applies: we identified the flows
\begin{equation}
(r,s)\to(|r-s+2|+1,1)\oplus(|r-s+2|+3,1)\oplus\dots\oplus(\textstyle{\min\binom{s+r-3}{2p-(s+r)+1}},1)\,.
\label{rule2a}
\end{equation}
In the Neveu--Schwarz case these flows, just like their ancestors,
change the type of the boundary conditions. The comment given in case
\ref{comment} is true here as well.
\item For \(s=2\)
\begin{equation}
(1,2)_\text{R}\longrightarrow(1,1)_\text{NS}\oplus(1,1)_{\widetilde{\mathrm{NS}}}\,.
\end{equation}
This is another example for a flow starting from a Ramond boundary
condition running to a Neveu--Schwarz one. The disorder line rule is
consistent with the flows
\begin{subequations}
\begin{align}
(r,2)_\text{NS}&\longrightarrow(r,1)_\text{R}\,,\\
(r,2)_{\widetilde{\mathrm{NS}}}&\longrightarrow(r,1)_\text{R}\,,\\
(r,2)_{\text{R}\;\,}&\longrightarrow(r,1)_\text{NS}\oplus(r,1)_{\widetilde{\mathrm{NS}}}\,.
\end{align}
\label{rule2b}
\end{subequations}
\end{enumerate}
\end{enumerate}

It is interesting that despite the lack of a complete classification
of consistent boundary states for \(p\) even, our results indicate the
same rules for these cases. The only exception is the boundary
condition corresponding to the supersymmetric
\((\frac{p}2,\frac{p+2}2)\) representation which could not be
investigated directly by TCSA.

It is also interesting to note that via the symmetry transformation of
the Kac table (\(r\to p-r\), \(s\to p+2-s\)) the rules \eqref{rule1a},
\eqref{rule1b} are mapped to the rules \eqref{rule2a} and
\eqref{rule2b}, respectively. This means that the two groups of rules
(\(\kappa<0\), \(\kappa>0\)) are not independent.

We note that although one can only approach and never reach the fixed
point, the endpoints are \((r,1)\) type boundary conditions that have
no relevant supersymmetric perturbations, so it is reasonable to take
our observations to be strong indications for the real result.

\end{section}

\begin{section}{Conclusions}
In this work we investigated the renormalisation group flows of
boundary conditions in \(N=1\) superconformal minimal models. Starting
from a pure Cardy-type boundary condition, the (supersymmetry
preserving) perturbating operator \(\hat G_{-1/2}\Phi_{1,3}\) induces
a renormalisation group flow whose infrared fixed points are again
superconformal invariant boundary conditions. As we have shown, the
Fredenhagen rules \cite{fredenhagen} for boundary flows in general
coset theories predict the IR fixed points of our flows. The predicted
flows obey a product rule called disorder line rule \cite{disorder}.

In this paper we used the TCSA method to check the Fredenhagen rules.
We found that these rules held in every case that we investigated. It
is interesting to note that there are flows which go from a
Neveu--Schwarz type boundary condition to a Ramond type, and vice
versa. Finally, we showed that all the flows obey the disorder line
rule.

Although for \(p\) even both these predictions and the description of
boundary states are flawed by the presence of the supersymmetric
representation, our analysis implies that a natural extension of the
rules are valid for these models.

There are several possibilities for continuing this work. One of them
is its extension to the study of the more complicated space of flows
of non-Cardy type boundary conditions. Another possibility is to
consider \(N=2\) models, which have direct relevance in string
theory. However, although the TCSA analysis should be straightforward
in principle, we expect too many states for reasonably high cuts,
which can spoil the applicability of TCSA in practice. It would be
interesting to include non-unitary models, too. The energies typically
become complex in these models (like for ordinary minimal models, as
reported in \cite{doreypocklington...}), which makes the
identification of the endpoint rather difficult.

\subsection*{Acknowledgement}
I would like to thank my supervisor, G\'abor Tak\'acs for his
guidance, help and the interesting discussions. I am also grateful to
Stefan Fredenhagen for his valuable remarks.

\end{section}

\appendix 

\section{Appendix}
In the appendix we present the results of the TCSA for the first
unitary models in detail. In the subsections we follow the
classification of the flows given in section \ref{class}. In the
tables the ``level'' entry indicates the level up to which the
agreement between the character and the calculated degeneracy pattern
in the IR holds. In the fourth field of every row the dimension of the
truncated Hilbert space is shown. The last field contains the number
of the corresponding figure.

\begin{figure}[h!]
\centering
\subfigure[Flows starting from b.c.\ \((1,3)\) in SM(3,5)]{\resizebox{74mm}{!}{\includegraphics{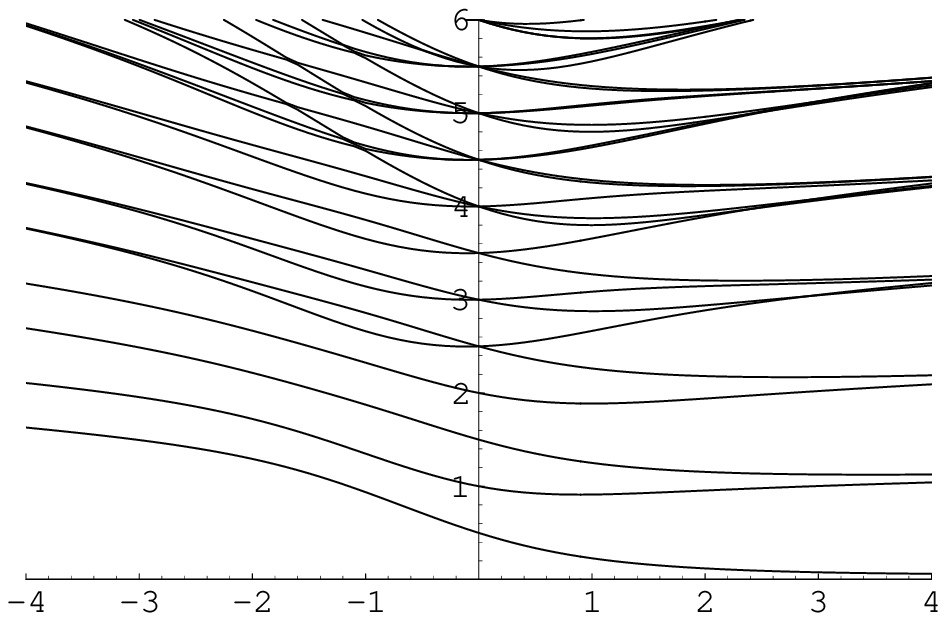}}\label{unnorm}}
\subfigure[Flows starting from b.c.\ \((1,3)\) in
  SM(3,5)]{\resizebox{74mm}{!}{\includegraphics{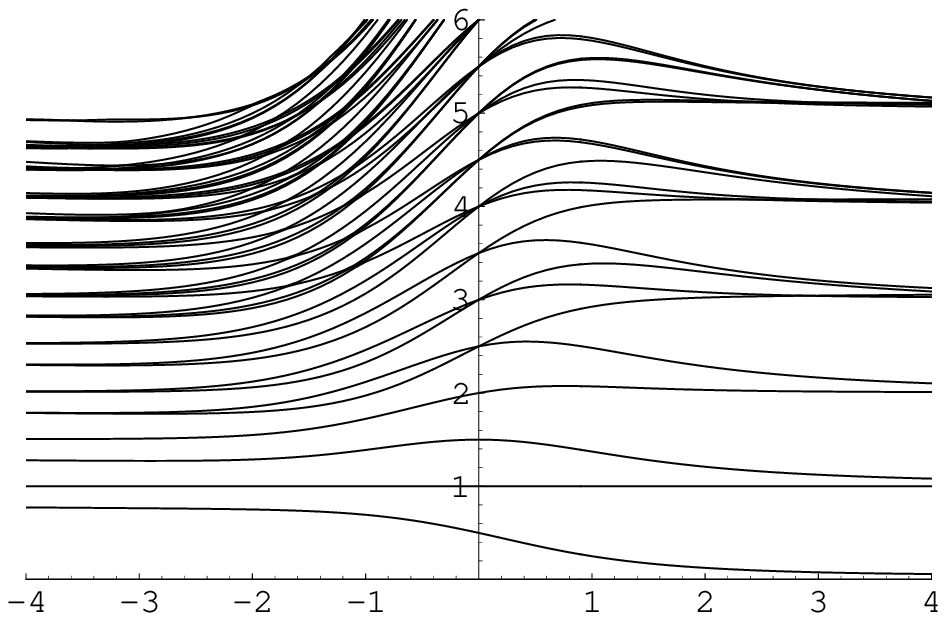}}\label{3,5,1,3}}
\end{figure}

In the figures collected at the end of the Appendix we plotted the
energy levels against the dimensionless coupling parameter
\(\kappa\). In these figures the starting boundary condition is
unprojected for the Neveu--Schwarz ones (i.e.\
\(\text{NS}\oplus\widetilde{\mathrm{NS}}\)), so these flows go to
unprojected NS or R boundary conditions. In figure \ref{unnorm} the
energy itself is plotted, while in all the other figures we plotted
the normalised energy
\((\varepsilon-\varepsilon_0)/(\varepsilon_i-\varepsilon_0)\) for some
appropriate \(i\) (usually 1).

\newpage

\begin{subsection}{First group of flows: \(\kappa>0\)}

\begin{subsubsection}{Flows starting from b.\ c.\ \((r,s)\) when \(s\le p-1\)}
These flows are predicted by the Fredenhagen rules and our results
agree with these predictions.
\[(r,s)\to(|s-r|+1,1)\oplus(|s-r|+3,1)\oplus\dots\oplus(\textstyle{\min\binom{s+r-1}{2p-(s+r)-1}},1)\]

\begin{center}
\begin{tabular}{|c|l|c|c|c|}
\hline
model&\hspace{0.9cm}flow&level&dim&fig.\\
\hline
\hline
SM\((3,5)\)&\((1,2)\to(2,1)\)&11&302&\ref{3,5,1,2}\\
\hline
SM\((4,6)\)&\((1,3)\to(3,1)\)&16&454&\ref{4,6,1,3}\\
\hline
&\((2,2)\to(1,1)\oplus(3,1)\)&17,15&536&\ref{4,6,2,2}\\
\hline
&\((1,2)\to(2,1)\)&11&414&\\
\hline
SM\((5,7)\)&\((1,3)\to(3,1)\)&13&388&\\
\hline
&\((2,2)\to(1,1)\oplus(3,1)\)&13,11&311&\\
\hline
&\((2,4)\to(3,1)\)&12&601&\ref{5,7,2,4}\\
\hline
&\((1,2)\to(2,1)\)&11&341&\\
\hline
&\((1,4)\to(4,1)\)&7&357&\ref{5,7,1,4}\\
\hline
&\((2,3)\to(2,1)\oplus(4,1)\)&8,6&303&\ref{5,7,2,3}\\
\hline
SM\((6,8)\)&\((1,3)\to(3,1)\)&14&396&\\
\hline
&\((1,5)\to(5,1)\)&8&540&\ref{6,8,1,5}\\
\hline
&\((2,2)\to(1,1)\oplus(3,1)\)&13,11&316&\\
\hline
&\((2,4)\to(3,1)\oplus(5,1)\)&12,8&440&\ref{6,8,2,4}\\
\hline
&\((1,2)\to(2,1)\)&10&344&\\
\hline
&\((1,4)\to(4,1)\)&7&385&\\
\hline
&\((2,3)\to(2,1)\oplus(4,1)\)&8,6&318&\\
\hline
&\((2,5)\to(4,1)\)&6&367&\ref{6,8,2,5}\\
\hline
&\((3,2)\to(2,1)\oplus(4,1)\)&9,8&317&\ref{6,8,3,2}\\
\hline
\end{tabular}
\end{center}

\end{subsubsection}

\begin{subsubsection}{Flows starting from b.\ c.\ \((r,s)\) when \(s=p\)}
These flows are not predicted by the Fredenhagen rules. Based on the
TCSA data we propose:
\[(r,p)\longrightarrow(p-r,1)\]

\begin{center}
\begin{tabular}{|c|l|c|c|c|}
\hline
model&\hspace{0.9cm}flow&level&dim&fig.\\
\hline
\hline
SM\((3,5)\)&\((1,3)\to(2,1)\)&6&407&\ref{3,5,1,3}\\
\hline
SM\((4,6)\)&\((1,4)\to(3,1)\)&18&490&\ref{4,6,1,4}\\
\hline
SM\((5,7)\)&\((1,5)\to(4,1)\)&7&298&\ref{5,7,1,5}\\
\hline
SM\((6,8)\)&\((1,6)\to(5,1)\)&14&289&\ref{6,8,1,6}\\
\hline
&\((2,6)\to(4,1)\)&7&426&\ref{6,8,2,6}\\
\hline
\end{tabular}
\end{center}

\end{subsubsection}
\end{subsection}

\begin{subsection}{Second group of flows: \(\kappa<0\)}

\begin{subsubsection}{Flows starting from b.\ c.\ \((r,s)\) when \(s\ge3\)}
These flows are again covered by the Fredenhagen rules. The TCSA
results verified these rules.
\[(r,s)\to(|r-s+2|+1,1)\oplus(|r-s+2|+3,1)\oplus\dots\oplus(\textstyle{\min\binom{s+r-3}{2p-(s+r)+1}},1)\]

\begin{center}
\begin{tabular}{|c|l|c|c|c|}
\hline
model&\hspace{0.9cm}flow&level&dim&fig.\\
\hline
\hline
SM\((3,5)\)&\((1,3)\to(1,1)\)&12&407&\ref{3,5,1,3}\\
\hline
SM\((4,6)\)&\((1,3)\to(1,1)\)&18&454&\ref{4,6,1,3}\\
\hline
&\((1,4)\to(2,1)\)&10&490&\ref{4,6,1,4}\\
\hline
SM\((5,7)\)&\((1,3)\to(1,1)\)&12&388&\\
\hline
&\((1,5)\to(3,1)\)&14&298&\ref{5,7,1,5}\\
\hline
&\((2,4)\to(1,1)\oplus(3,1)\)&15,13&601&\ref{5,7,2,4}\\
\hline
&\((1,4)\to(2,1)\)&9&357&\ref{5,7,1,4}\\
\hline
&\((2,3)\to(2,1)\)&8&303&\ref{5,7,2,3}\\
\hline
SM\((6,8)\)&\((1,3)\to(1,1)\)&14&396&\\
\hline
&\((1,5)\to(3,1)\)&14&540&\ref{6,8,1,5}\\
\hline
&\((2,4)\to(1,1)\oplus(3,1)\)&13,11&440&\ref{6,8,2,4}\\
\hline
&\((2,6)\to(3,1)\oplus(5,1)\)&17,13&426&\ref{6,8,2,6}\\
\hline
&\((1,6)\to(4,1)\)&9&289&\ref{6,8,1,6}\\
\hline
&\((1,4)\to(2,1)\)&9&385&\\
\hline
&\((2,3)\to(2,1)\)&7&318&\\
\hline
&\((2,5)\to(2,1)\oplus(4,1)\)&7,5&367&\ref{6,8,2,5}\\
\hline
\end{tabular}
\end{center}

\end{subsubsection}

\begin{subsubsection}{Flows starting from b.\ c.\ \((r,s)\) when \(s=2\)}
Finally, we present the second type of exceptional (unpredicted)
flows:
\[(r,2)\longrightarrow(r,1)\]

\begin{center}
\begin{tabular}{|c|l|c|c|c|}
\hline
model&\hspace{0.9cm}flow&level&dim&fig.\\
\hline
\hline
SM\((3,5)\)&\((1,2)\to(1,1)\)&9&302&\ref{3,5,1,2}\\
\hline
SM\((4,6)\)&\((2,2)\to(2,1)\)&9&536&\ref{4,6,2,2}\\
\hline
&\((1,2)\to(1,1)\)&22&414&\\
\hline
SM\((5,7)\)&\((2,2)\to(2,1)\)&7&311&\\
\hline
&\((1,2)\to(1,1)\)&20&341&\\
\hline
SM\((6,8)\)&\((2,2)\to(2,1)\)&7&316&\\
\hline
&\((1,2)\to(1,1)\)&20&344&\\
\hline
&\((3,2)\to(3,1)\)&16&317&\ref{6,8,3,2}\\
\hline
\end{tabular}
\end{center}

\end{subsubsection}

\end{subsection}

\newpage

\section*{Figures}

\begin{figure}[h!]
\centering
\subfigure[Flows starting from b.c.\ \((1,2)\) in
  SM(3,5)]{\resizebox{74mm}{!}{\includegraphics{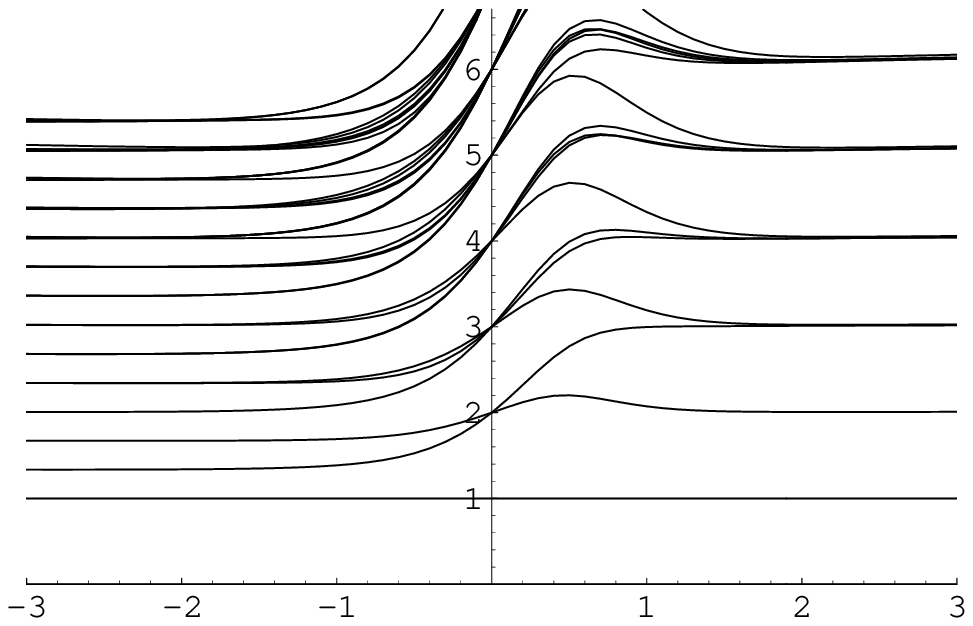}}\label{3,5,1,2}}
\caption{Flows in SM(3,5)}
\end{figure}

\clearpage

\begin{figure}[h!]
\centering
\subfigure[Flows starting from b.c.\ \((1,3)\) in SM(4,6)]{\resizebox{74mm}{!}{\includegraphics{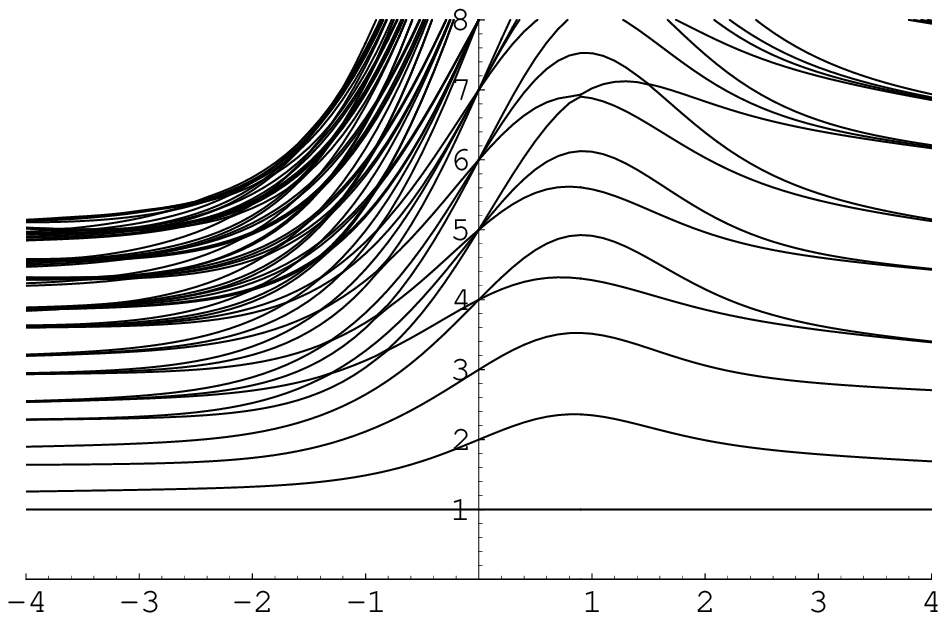}}\label{4,6,1,3}}
\subfigure[Flows starting from b.c.\ \((2,2)\) in
  SM(4,6)]{\resizebox{74mm}{!}{\includegraphics{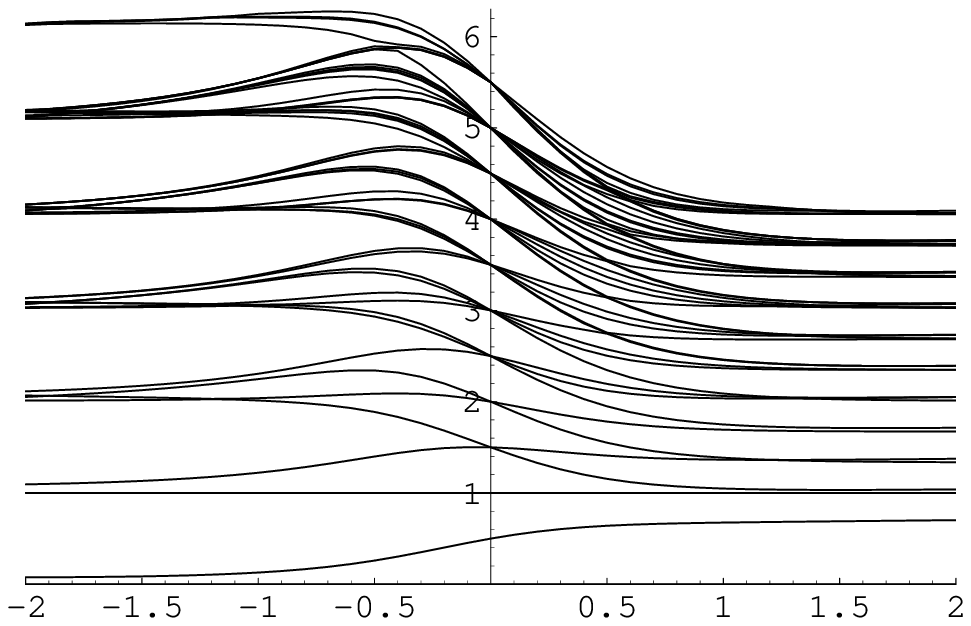}}\label{4,6,2,2}}
\end{figure}

\begin{figure}[h!]
\centering
\subfigure[Flows starting from b.c.\ \((1,4)\) in
  SM(4,6)]{\resizebox{74mm}{!}{\includegraphics{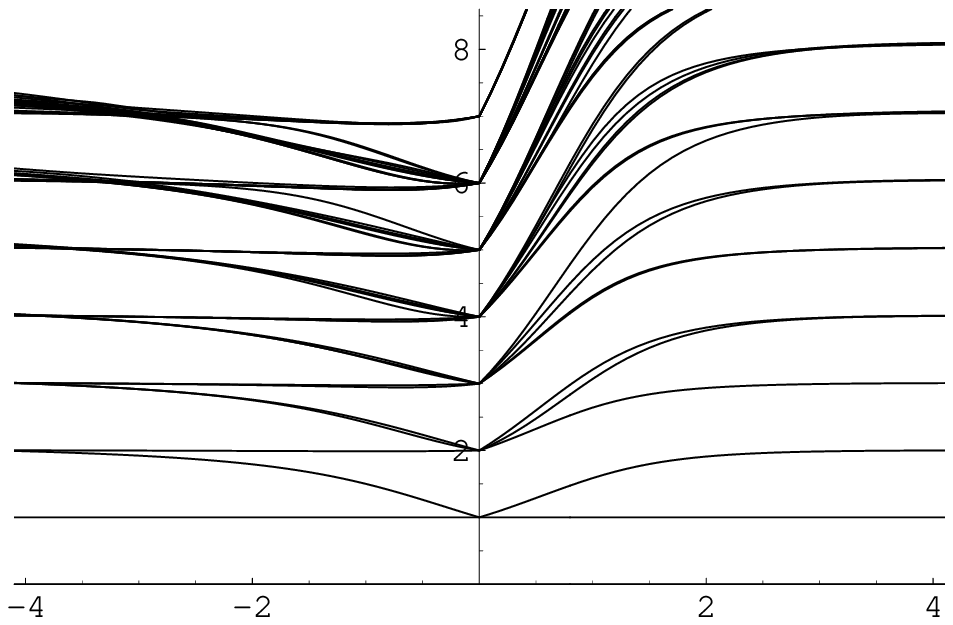}}\label{4,6,1,4}}
\caption{Flows in SM(4,6)}
\end{figure}

\clearpage

\begin{figure}[h!]
\centering
\subfigure[Flows starting from b.c.\ \((1,5)\) in
  SM(5,7)]{\resizebox{74mm}{!}{\includegraphics{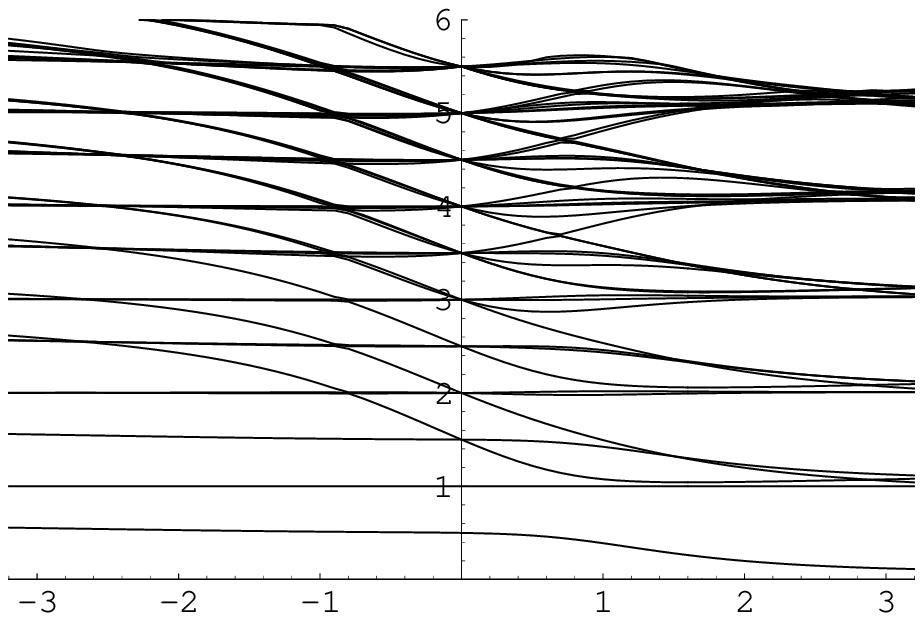}}\label{5,7,1,5}}
\subfigure[Flows starting from b.c.\ \((2,4)\) in
  SM(5,7)]{\resizebox{74mm}{!}{\includegraphics{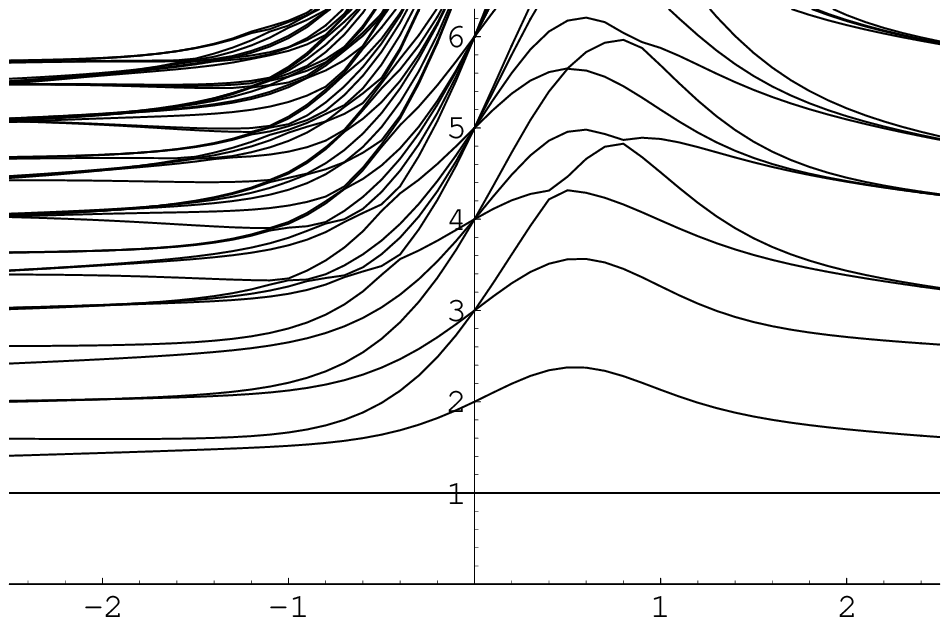}}\label{5,7,2,4}}
\end{figure}

\begin{figure}[h!]
\centering
\subfigure[Flows starting from b.c.\ \((1,4)\) in SM(5,7)]{\resizebox{74mm}{!}{\includegraphics{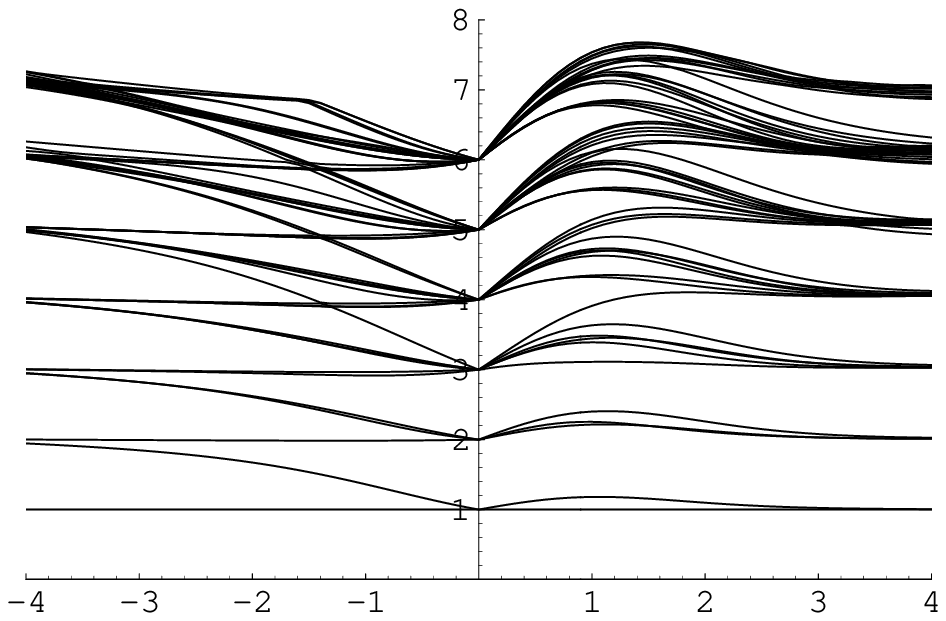}}\label{5,7,1,4}}
\subfigure[Flows starting from b.c.\ \((2,3)\) in
  SM(5,7)]{\resizebox{74mm}{!}{\includegraphics{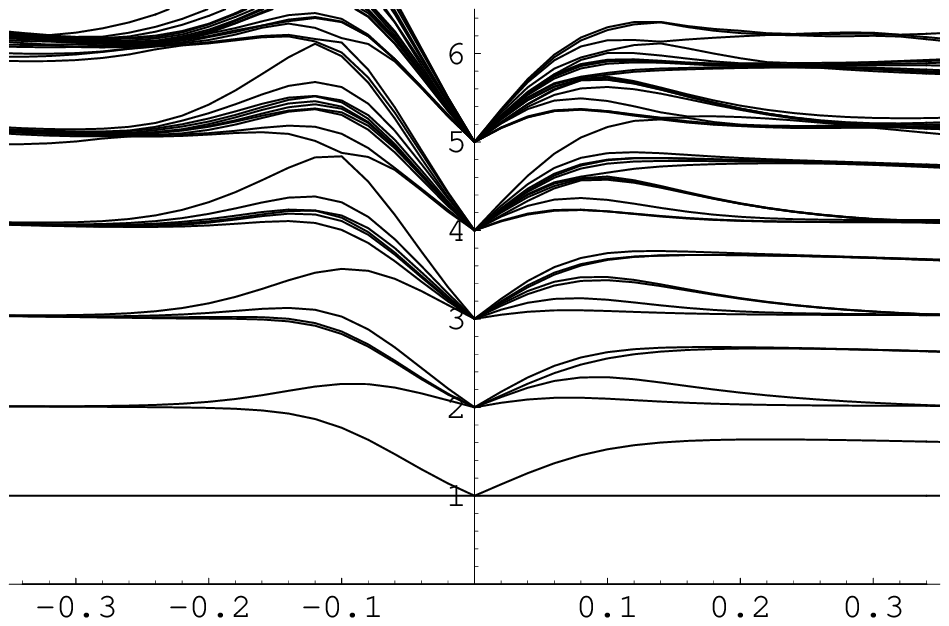}}\label{5,7,2,3}}
\caption{Flows in SM(5,7)}
\end{figure}

\clearpage

\begin{figure}[h!]
\centering
\subfigure[Flows starting from b.c.\ \((1,5)\) in SM(6,8)]{\resizebox{74mm}{!}{\includegraphics{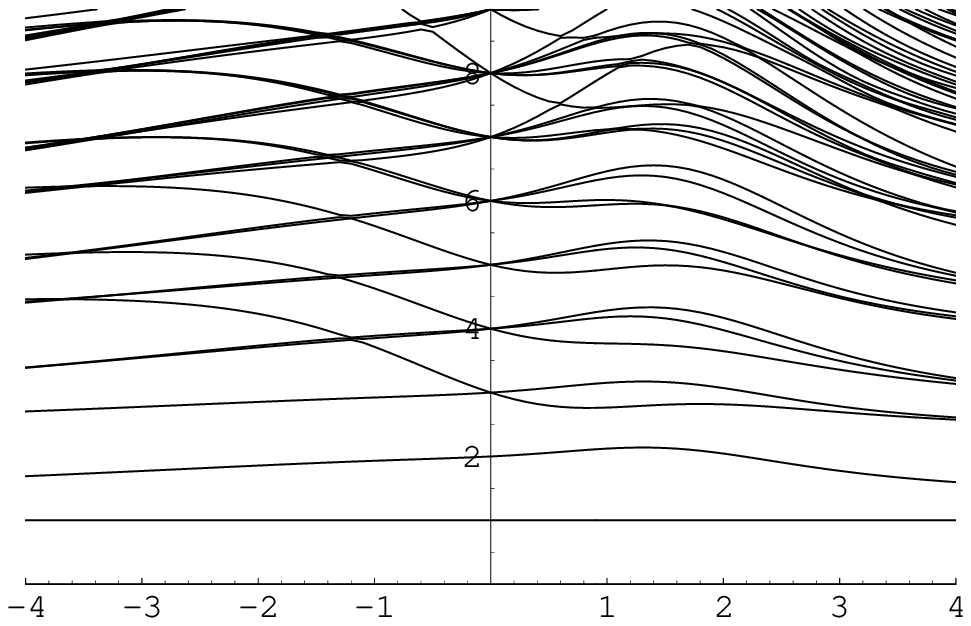}}\label{6,8,1,5}}
\subfigure[Flows starting from b.c.\ \((2,4)\) in SM(6,8)]{\resizebox{74mm}{!}{\includegraphics{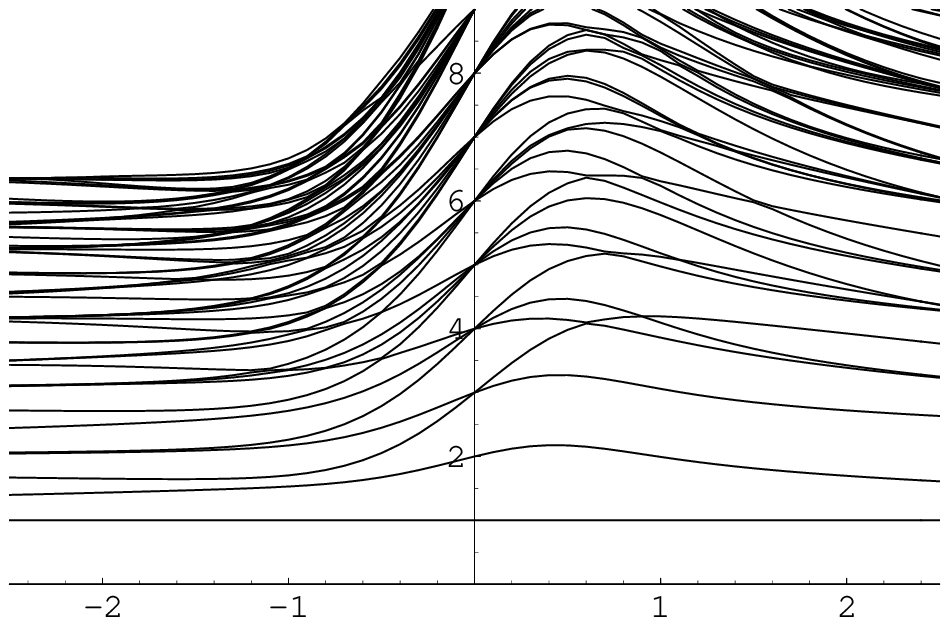}}\label{6,8,2,4}}
\end{figure}

\begin{figure}[h!]
\centering 
\subfigure[Flows starting from b.c.\ \((2,6)\) in SM(6,8)]{\resizebox{74mm}{!}{\includegraphics{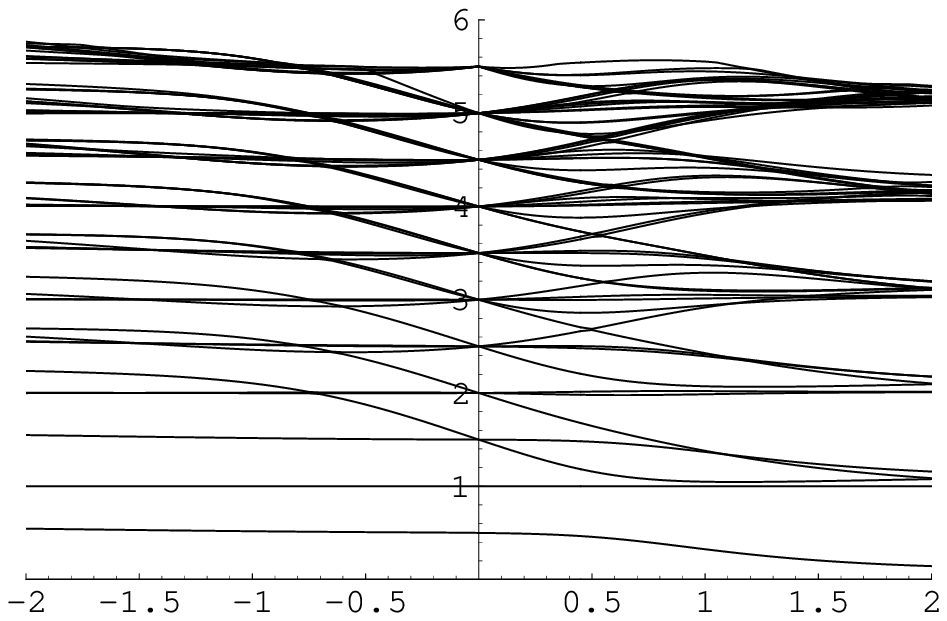}}\label{6,8,2,6}}
\subfigure[Flows starting from b.c.\ \((2,5)\) in
SM(6,8)]{\resizebox{74mm}{!}{\includegraphics{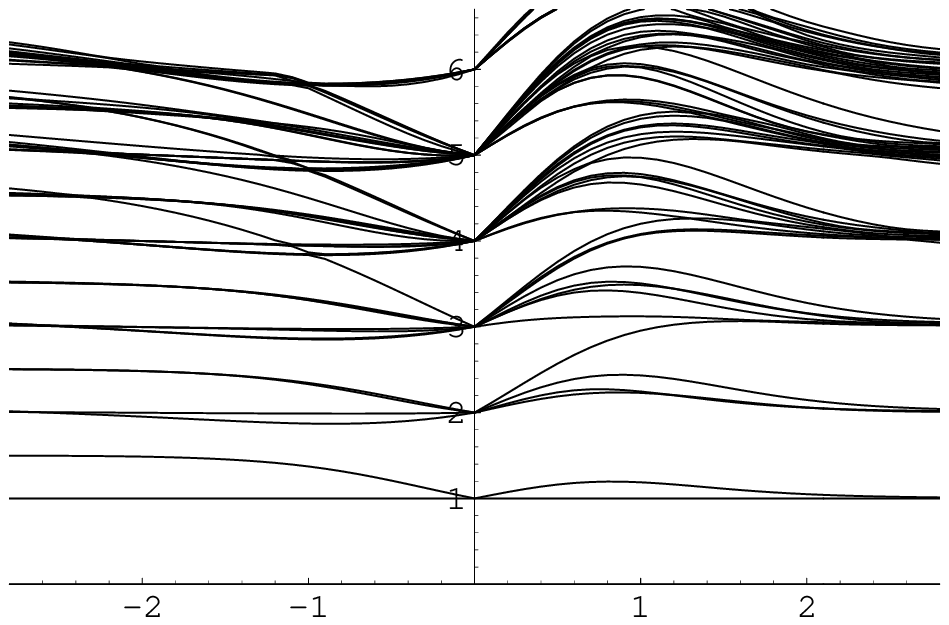}}\label{6,8,2,5}}
\end{figure}

\begin{figure}[h!]
\centering
\subfigure[Flows starting from b.c.\ \((3,2)\) in
  SM(6,8)]{\resizebox{74mm}{!}{\includegraphics{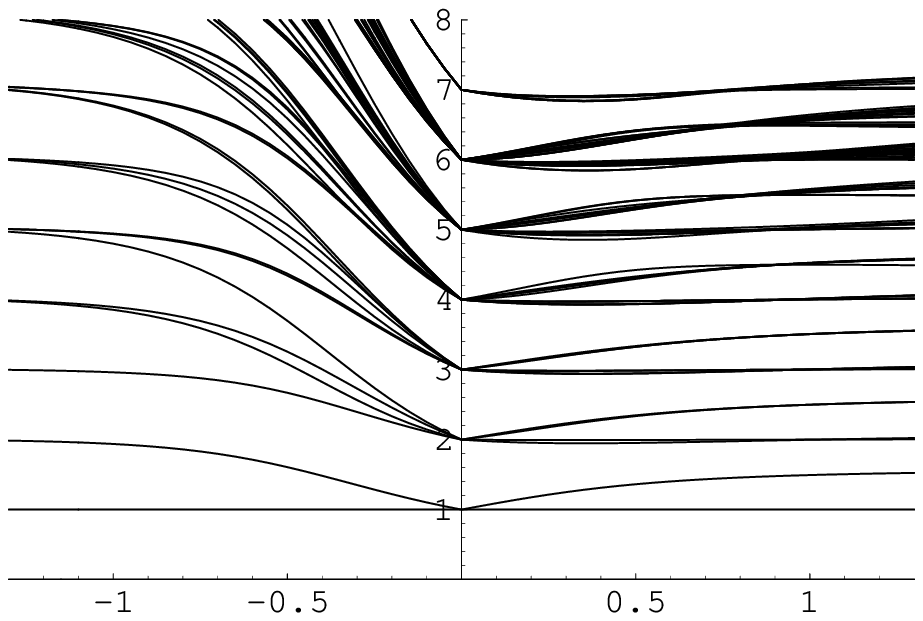}}\label{6,8,3,2}}
\subfigure[Flows starting from b.c.\ \((1,6)\) in
  SM(6,8)]{\resizebox{74mm}{!}{\includegraphics{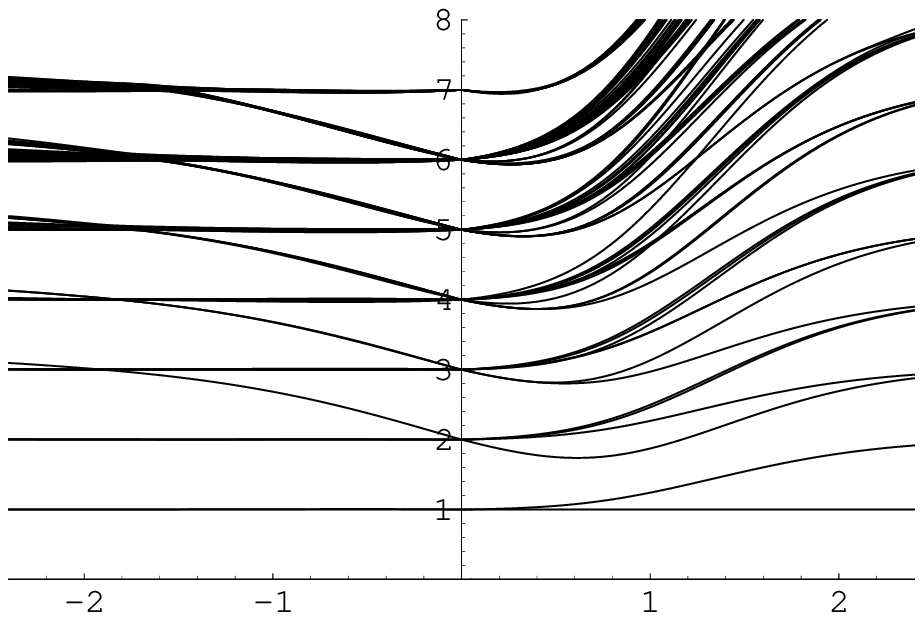}}\label{6,8,1,6}}
\caption{Flows in SM(6,8)}
\end{figure}


\newpage

\begin{thebibliography}{99}
\raggedright
\bibitem{saleur}H. Saleur: {\em Lectures on Non Perturbative Field
  Theory and Quantum Impurity Problems}, \texttt{cond-mat/9812110}
\bibitem{openstrings}C. Angelantonj, A. Sagnotti: {\em Open Strings},
  \textsl{Phys. Rept.} \textbf{371} (2002) 1 (\texttt{hep-th/0204089})
\bibitem{tachyons}J. A. Harvey, D. Kutasov, E. J. Martinec: {\em On
  the relevance of tachyons}, \texttt{hep-th/0003101}
\bibitem{reckrogg}A. Recknagel, D. Roggenkamp, V. Schomerus: {\em On
  relevant boundary perturbations of unitary minimal models}, \textsl{Nucl. Phys.} \textbf{B558} (2000) 552 (\texttt{hep-th/0003110})
\bibitem{gfunc}P. Dorey, I. Runkel, R. Tateo, G. Watts: {\em
  g-function flow in perturbed boundary conformal field theories},
  \textsl{Nucl. Phys.} \textbf{B578} (2000) 85
  (\texttt{hep-th/9909216})
\bibitem{lesage}F. Lesage, H. Saleur, P. Simonetti: {\em Boundary
  flows in minimal models}, \textsl{Phys. Lett.} \textbf{B427} (1998)
  85 (\texttt{hep-th/9802061})
\bibitem{c=1}K. Graham, I. Runkel, G. M. T. Watts: {\em Minimal model
  boundary flows and \(c=1\) CFT}, \textsl{Nucl. Phys.} \textbf{B608}
  (2001) 527 (\texttt{hep-th/0101187})
\bibitem{grahamrunkelwatts}K. Graham, I. Runkel, G. M. T. Watts:
  {\em Renormalisation group flows of boundary theories}, (\texttt{hep-th/0010082})
\bibitem{yurovzam}V. P. Yurov, Al. B. Zamolodhikov: {\em Truncated
  conformal space approach to the scaling Lee-Yang model},
  \textsl{Int. J. Mod. Phys.} \textbf{A5} (1990) 3221
\bibitem{doreypocklington...}P. Dorey, A. Pocklington, R. Tateo,
  G. Watts: {\em TBA and TCSA with boundaries and excited states},
  \textsl{Nucl. Phys.}  \textbf{B525} (1998) 641
  (\texttt{hep-th/9712197}) 
\bibitem{watts}G. M. T. Watts: {\em Null vectors of the superconfomal
  algebra: the Ramond sector}, \textsl{Nucl.\ Phys.} \textbf{B407}
(1993) 213 (\texttt{hep-th/9306034})
\bibitem{nepomechie}R. I. Nepomechie: {\em Consistent superconformal
  boundary states}, \textsl{J. Phys.} \textbf{A34}
  (2001) 6509 (\texttt{hep-th/0102010})
\bibitem{ishibashi}N. Ishibashi: {\em  The Boundary and Crosscap
  States in Conformal Field Theory}, \textsl{Mod. Phys. Lett}
  \textbf{A4} (1989) 251
\bibitem{cardy}J. L. Cardy: {\em Boundary conditions, fusion rules and
  the Verlinde formula}, \textsl{Nucl. Phys.} \textbf{B324} (1989) 581
\bibitem{mathieu}P. Mathieu: {\em Integrability of perturbed superconformal
  minimal models}, \textsl{Nucl. Phys.} \textbf{B336} (1990) 338
\bibitem{fredenhagen} S. Fredenhagen: {\em Organizing boundary RG
  flows}, \textsl{Nucl. Phys.} \textbf{B660} (2003) 436
  (\texttt{hep-th/0301229})
\bibitem{gko} P. Goddard, A. Kent, D. I. Olive: {\em Unitary
  representations of the Virasoro and super-Virasoro algebras},
  \textsl{Commun.\ Math.\ Phys.} \textbf{103} (1986) 105
\bibitem{disorder} K. Graham, G. M. T. Watts: {\em Defect Lines and
  Boundary Flows}, \textsl{JHEP} \textbf{0404} (2004) 019 (\texttt{hep-th/0306167})
\bibitem{ahn-rim} Ch. Ahn, Ch. Rim: {\em Boundary Flows in general Coset
  Theories}, \textsl{J. Phys.} \textbf{A32} (1999) 2509 (\texttt{hep-th/9805101})
\end{thebibliography}
\end{document}